\documentclass[10pt,aps,prb,twocolumn,amsmath,amssymb,
longbibliography,
superscriptaddress
]{revtex4-2}

\usepackage{dcolumn}
\usepackage{bm}
\usepackage[normalem]{ulem}
\usepackage[colorlinks=true,linkcolor=blue,citecolor=blue,urlcolor=blue,breaklinks=true]{hyperref}
\usepackage{verbatim}
\usepackage[ruled,vlined]{algorithm2e}
\usepackage{tikz}
\usepackage{braket}
\usepackage{listings}
\usepackage{amsfonts}
\usepackage{amsmath}
\usepackage{amssymb}
\usepackage{amsbsy} %for sigma in bold face
\usepackage{graphicx}
\usepackage{appendix}
\usepackage{color}
\usepackage{amsthm}
\newtheorem{proposition}{Proposition}
\newtheorem{remark}{Remark}
\usepackage{mathrsfs}

\usepackage{mathdots}
\newtheorem{lemma}{Lemma}
\newtheorem{theorem}{Theorem}

\newcommand{\open}{\mathrm{open}}

\newcommand{\cur}{\mathrm{cur}}
\newcommand{\en}{\mathrm{en}}
\newcommand{\esc}{\mathrm{esc}}

\newcommand{\avg}[1]{\overline{#1}}
\newcommand{\free}{\mathrm{free}}
\newcommand{\cut}{\mathrm{cut}}
\newcommand{\Ran}{\mathrm{Ran}}
\newcommand{\decay}{{decay}}
\newcommand{\prop}{\mathrm{prop}}

\newcommand{\conn}{\mathrm{conn}}
\newcommand{\ev}{\mathrm{ev}}
\newcommand{\edge}{\mathrm{edge}}
\newcommand{\diag}{\mathbf{diag}}
\newcommand{\sgn}{\mathbf{sgn}}

\begin{document}

\title{Scattering-state theory of open Floquet lattices: transfer matrices, branch openness, and robust asymmetry}

\author{Ren Zhang} \email{physiren@outlook.com}
 \affiliation{New Cornerstone Science Laboratory, Department of Physics,\\ \text{The Hong Kong University of Science and Technology,
Clear Water Bay, Kowloon 999077, Hong Kong, China}}

\author{Xiao-Yu Ouyang}
 \affiliation{New Cornerstone Science Laboratory, Department of Physics,\\ \text{The Hong Kong University of Science and Technology,
Clear Water Bay, Kowloon 999077, Hong Kong, China}}
 \affiliation{\text{Division of Chemistry and Chemical Engineering, California Institute of Technology, Pasadena, CA 91125, USA}}

\author{Xu-Dong Dai}
 \affiliation{New Cornerstone Science Laboratory, Department of Physics,\\ \text{The Hong Kong University of Science and Technology,
Clear Water Bay, Kowloon 999077, Hong Kong, China}}

\author{Xi Dai} \email{ daix@ust.hk }
\affiliation{New Cornerstone Science Laboratory, Department of Physics,\\ \text{The Hong Kong University of Science and Technology,
Clear Water Bay, Kowloon 999077, Hong Kong, China}}

\begin{abstract}
We establish a scattering-state theory for open one-dimensional Floquet lattices based on a frequency-domain transfer-matrix formulation. For real quasienergy, the conjugate-symplectic structure of the transfer matrix separates bulk Floquet--Bloch modes into propagating and evanescent sectors, enabling a consistent treatment of interface matching and the shrinking-window smoothing required for long-sample transport. By tracking how incoming states populate deep-bulk propagating branches, we define branch-resolved weights \(p_{\mu\alpha}\) and total branch weights \(p_\mu\). We prove that \(p_\mu\) equals the escape probability of a wave packet initialized on the corresponding branch. In the open geometries considered here, true bound trapping of propagating branches is nongeneric, yielding \(p_\mu=1\) for generic parameters. This generic openness implies that long-sample transport is governed by deep-bulk branch populations rather than by boundary-sensitive interference. Consequently, the integrated left--right transmission asymmetry reduces to the net chirality, and hence the winding contribution, of an isolated Floquet band. The robust topological observable is therefore the accumulated asymmetry plateau, not the detailed transmission line shape, which remains strongly reshaped by nonadiabatic boundaries. A spatially adiabatic boundary serves only as a transparent benchmark for resolving the branch structure, not as the origin of the topological response.
\end{abstract}

\maketitle

\section{Introduction}
\label{sec:intro}

Periodic driving qualitatively reorganizes the spectral and transport properties of quantum matter. In spatially periodic systems, the Floquet--Bloch framework combines temporal and spatial periodicities into quasienergy bands defined modulo \(\omega\), supporting phenomena with no static analogue, such as winding-based topology and quantized pumping \cite{Floquet1883,Shirley1965,Sambe1973,Thouless1983,Kitagawa2010,Rudner2013}. While this framework is well established for closed systems, transport through an \emph{open} Floquet lattice raises a more fundamental question: what is the correct scattering-state formulation for long samples, and which observable cleanly exposes the underlying bulk topology in the presence of realistic boundaries?

This problem presents two structural challenges. First, an open device requires asymptotic leads. When these leads are undriven, the interfaces to the driven lattice open multiple Floquet sideband channels, inducing substantial mode conversion and backscattering. Second, transport through an extended Floquet sample constitutes a long-sample problem: finite segments exhibit strong Fabry--P\'erot-type oscillations in their scattering amplitudes \cite{Born1999}. Consequently, the physically relevant observable is not the pointwise transmission coefficient at a fixed energy or quasienergy, but an appropriately coarse-grained long-sample quantity.

This paper presents the complete theoretical framework for the accompanying article \cite{Ljoint}, which establishes that in open Floquet lattices the robust observable is the integrated left--right transmission asymmetry rather than the detailed transmission line shape. Here we construct the general scattering-state theory that formalizes this result. We begin with a transfer-matrix formulation in frequency space. For real quasienergy, the spatial first-order Floquet evolution preserves a bilinear current form, rendering the transfer matrix conjugate-symplectic \cite{Feng2010,Weyl1939,Pendry1994,Haber2017}. This algebraic structure separates bulk modes into propagating and evanescent sectors and provides a natural framework for interface matching, finite-length composition, and the shrinking-window smoothing required to define long-sample transport observables.

Building on this structure, we formulate the open sector through branch-resolved scattering states. Rather than describing openness solely via asymptotic lead channels, we determine how incoming scattering states populate propagating Floquet--Bloch branches deep inside a long sample. This yields branch-resolved weights \(p_{\mu\alpha}\) and total branch weights \(p_\mu\). We prove that \(p_\mu\) is exactly equal to the escape probability of a wave packet initialized on the corresponding bulk branch. This identity cleanly separates two distinct physical questions: which bulk branches are accessed by incident states, and whether those branches remain genuinely open.

A central result is the generic-openness principle: for propagating branches, permanent trapping would require a genuine Floquet bound state. In the open geometries considered here, such trapping is nongeneric because it demands an overdetermined matching between decaying and propagating sectors across the sample. For generic parameter values, therefore, \(p_\mu=1\). This generic openness converts branch-resolved accessibility directly into a topological transport statement.

The boundary-robust transmission asymmetry identified in the companion article follows directly from this structure. After shrinking-window smoothing, transport depends on deep-bulk branch populations rather than on boundary-sensitive interference. When an incident-energy window selectively populates an isolated Floquet band, the integrated left--right transmission asymmetry reduces to the net chirality of that band, and thus to its winding contribution. The robust topological signature is therefore not the detailed transmission line shape---which is strongly reshaped by nonadiabatic boundaries---but the accumulated asymmetry plateau. Within this theory, a spatially adiabatic boundary serves only as a transparent benchmark: it minimizes branch mixing and clarifies the spectroscopic realization of the branch structure, but it does not generate the topological response.

The remainder of the paper is organized as follows. Section~\ref{sec:tm_csp} develops the transfer-matrix formulation and establishes its conjugate-symplectic structure. Section~\ref{sec:interface_scattering} constructs interface scattering, finite-length composition, and the smoothed transmittance. Section~\ref{sec:open_sector} introduces the branch-resolved open sector and proves the equivalence between long-sample branch weights and escape probabilities. 
Section~\ref{sec:generic_openness} demonstrates that true bound-state trapping of propagating branches is nongeneric, establishing \(p_\mu=1\) for generic parameter values. 
Section~\ref{sec:robust_asymmetry} derives the boundary-robust long-sample transmission asymmetry and examines the adiabatic boundary as a clean benchmark. Section~\ref{sec:conclusion} summarizes our findings.

\section{Transfer-matrix formulation and conjugate-symplectic structure}
\label{sec:tm_csp}

In this section we develop the structural foundation of the scattering state theory.
The basic logic is as follows.
We first rewrite the time-periodic Schr\"odinger equation as a first-order evolution problem in the spatial coordinate.
The resulting transfer matrix carries a conserved bilinear current form and is therefore conjugate-symplectic.
This structure organizes the modes in Floquet lattice into propagating and evanescent sectors, enforces reciprocal-conjugate spectral pairing, and yields the \(J\)-adapted dual basis needed later for interface matching and finite-length composition.
Some proofs are short enough to keep implicit in the main line, while the algebraic proofs that would otherwise interrupt the transport logic are deferred to Appendices~\ref{app:tm_algebraic} and \ref{app:tm_dualnorm}.

Throughout the paper we use
\[
d=1,\qquad m=\frac12,\qquad \hbar=1.
\]
where \(d\) is lattice constant of Floquet lattice and \(m\) as particle mass.

\subsection{Frequency-domain first-order system}
\label{subsec:tm_first_order}

We start from the time-periodic Schr\"odinger equation
\begin{equation}
\begin{aligned}
i\partial_t\psi(x,t) &= \left[-\partial_x^2+V(x,t)\right]\psi(x,t), \\
V(x,t+\tau) &= V(x,t), \qquad \tau=\frac{2\pi}{\omega}.
\end{aligned}
\label{eq:tm_sch}
\end{equation}
Using the Floquet ansatz
\begin{equation}
\psi(x,t)
=
e^{-i\epsilon t}\sum_{n\in\mathbb Z}\varphi_n(x,\epsilon)e^{-in\omega t},
\label{eq:tm_floquet_ansatz}
\end{equation}
one obtains the coupled second-order system \cite{Emmanouilidou2002,Wenjun1999,Martinez2001}
\begin{equation}
\begin{aligned}
\bm\varphi''(x,\epsilon) &= M(x,\epsilon)\bm\varphi(x,\epsilon), \\
M_{mn}(x,\epsilon) &= -(\epsilon+n\omega)\delta_{mn}+V_{m-n}(x).
\end{aligned}
\label{eq:tm_second_order}
\end{equation}
where
\[
\bm\varphi=(\ldots,\varphi_{-1},\varphi_0,\varphi_1,\ldots)^T.
\]

It is convenient to introduce the wave-function vector \(Y\) and the dynamic matrix \(G\), which also serves as the generator of the transfer matrix.
\begin{equation}
Y(x,\epsilon)
=
\begin{bmatrix}
\bm\varphi(x,\epsilon)\\[2pt]
\bm\varphi'(x,\epsilon)
\end{bmatrix},
\qquad
G(x,\epsilon)
=
\begin{bmatrix}
0&I\\
M(x,\epsilon)&0
\end{bmatrix}.
\label{eq:tm_wfv}
\end{equation}
The Floquet problem is then rewritten as a first-order spatial evolution,
\begin{equation}
\partial_x Y(x,\epsilon)=G(x,\epsilon)Y(x,\epsilon).
\label{eq:tm_first_order}
\end{equation}
The associated transfer matrix is
\begin{equation}
\begin{aligned}
F(x_2,x_1;\epsilon) &= \mathcal X \exp\!\left[ \int_{x_1}^{x_2} G(x,\epsilon)\,dx \right], \\
Y(x_2,\epsilon) &= F(x_2,x_1;\epsilon)\,Y(x_1,\epsilon).
\end{aligned}
\label{eq:tm_transfer}
\end{equation}
where \(\mathcal X\) denotes spatial ordering.

Equation~\eqref{eq:tm_first_order} is the starting point of the entire construction below.
All later channel decompositions, interface formulas, and long-sample asymptotics are formulated in terms of \(F\).

\subsection{Conserved current and conjugate-symplectic structure}
\label{subsec:tm_csp_current}

Define
\begin{equation}
J=
\begin{bmatrix}
0&I\\
-I&0
\end{bmatrix}.
\label{eq:tm_J}
\end{equation}
Since \(M(x,\epsilon)\) is Hermitian for real \(\epsilon\), the generator satisfies
\begin{equation}
G^\dagger(x,\epsilon)J+JG(x,\epsilon)=0.
\label{eq:tm_inf_csp}
\end{equation}
As a result, the transfer matrix obeys the finite conjugate-symplectic constraint
\begin{equation}
F(x_2,x_1;\epsilon)^\dagger J F(x_2,x_1;\epsilon)=J.
\label{eq:tm_csp}
\end{equation}

The time-averaged probability current takes the bilinear form
\begin{equation}
j(x,\epsilon)=-i\,Y(x,\epsilon)^\dagger JY(x,\epsilon),
\label{eq:tm_current}
\end{equation}
and Eq.~\eqref{eq:tm_csp} is precisely the statement that this current is conserved along the spatial evolution.

Equation~\eqref{eq:tm_csp} is the basic structural input for the scattering state theory.
It controls the organization of the bulk spectrum, the \(J\)-orthogonality relations between Floquet--Bloch branches, and the dual-basis algebra used later in channel matching.

\subsection{Modes in Floquet lattices}
\label{subsec:tm_bulk_modes}

In a homogeneous bulk region with spatial period \(1\), define the one-cell transfer matrix
\begin{equation}
F(\epsilon)\equiv F(1,0;\epsilon).
\label{eq:tm_unitcell_F}
\end{equation}
Modes are obtained from the eigenvalue problem
\begin{equation}
F(\epsilon)\,y_\mu(\epsilon)=\lambda_\mu(\epsilon)\,y_\mu(\epsilon).
\label{eq:tm_cell_eig}
\end{equation}
If \(|\lambda_\mu|=1\), we write
\begin{equation}
\lambda_\mu(\epsilon)=e^{ik_\mu(\epsilon)}
\label{eq:tm_lambda_k}
\end{equation}
and call the branch propagating.
If \(|\lambda_\mu|\neq1\), the branch is evanescent.

The corresponding bulk wave-function vector is
\begin{equation}
Y_\mu(x,\epsilon)=F(x,0;\epsilon)y_\mu(\epsilon),
\label{eq:tm_bulk_wfv}
\end{equation}
and its current
\begin{equation}
j_\mu(\epsilon)=-i\,Y_\mu(x,\epsilon)^\dagger JY_\mu(x,\epsilon)
\label{eq:tm_branch_current}
\end{equation}
is independent of \(x\).

Throughout, we call a quasienergy \(\epsilon\) \emph{regular} if the relevant bulk branches are simple, no propagating branch has vanishing current, and the one-cell transfer matrix \(F(\epsilon)\) is diagonalizable.
This is the regime in which channel assignments, reciprocal pairing, and current normalizations are unambiguous.
In the rest of the paper we always work at regular quasienergy unless explicitly stated otherwise.
Singular points such as band tops and band bottoms do occur physically, but they form a measure-zero set and do not affect the long-sample quantities studied here.

\subsection{Reciprocal-conjugate pairing, \(J\)-orthogonality, and sign equal division}
\label{subsec:tm_pairing_count}

We now discuss the spectral structure of the one-cell transfer matrix
\[
F(\epsilon)\in\mathbb C^{2N\times 2N},
\]
where \(N\) is the frequency truncation dimension.
At a regular quasienergy \(\epsilon\), we denote by \(N_B(\epsilon)\) the number of right-going propagating branches and by \(N_E(\epsilon)\) the number of right-decaying evanescent branches.
The corresponding left-going and left-decaying numbers will be shown below to be equal, and hence
\begin{equation}
N_B(\epsilon)+N_E(\epsilon)=N.
\label{eq:tm_NB_NE_sum}
\end{equation}

The conjugate-symplectic condition
\[
F^\dagger JF=J
\]
imposes two basic spectral constraints.

\begin{lemma}[Reciprocal-conjugate pairing]
\label{lem:tm_pairing}
If \(F^\dagger JF=J\) and \(\lambda\in\sigma(F)\), then
\begin{equation}
\frac{1}{\lambda^*}\in\sigma(F).
\label{eq:tm_reciprocal_pair}
\end{equation}
\end{lemma}

\begin{lemma}[\(J\)-orthogonality]
\label{lem:tm_Jorth}
Let \(Fy_m=\lambda_m y_m\) and \(Fy_n=\lambda_n y_n\).
If \(\lambda_m^*\lambda_n\neq1\), then
\begin{equation}
y_m^\dagger J y_n=0.
\label{eq:tm_Jorth}
\end{equation}
\end{lemma}

The proofs are given in Appendix~\ref{appsubsec:tm_pairing}.

It is therefore natural to define, for each eigenvector \(y_\mu\), a paired eigenvector \(\bar y_\mu\) by
\begin{equation}
F\bar y_\mu=\bar\lambda_\mu\,\bar y_\mu,
\qquad
\bar\lambda_\mu=\frac{1}{\lambda_\mu^*},
\label{eq:tm_pair_def}
\end{equation}
with the convention
\begin{equation}
\bar{\bar y}_\mu=y_\mu.
\label{eq:tm_pair_involution}
\end{equation}
If \(|\lambda_\mu|=1\), then \(\bar\lambda_\mu=\lambda_\mu\), so the branch is self-paired; in that case we may choose
\begin{equation}
\bar y_\mu=y_\mu.
\label{eq:tm_selfpair_choice}
\end{equation}
If \(|\lambda_\mu|\neq1\), then \(\bar y_\mu\) is a distinct reciprocal-conjugate partner.

\begin{figure}
    \centering
    \includegraphics[width=1\linewidth]{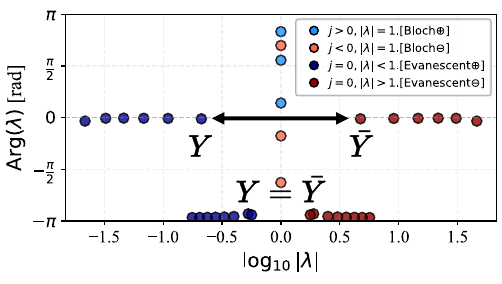}
    \caption{Schematic spectrum of a conjugate-symplectic transfer matrix. If \(\lambda\) is an eigenvalue, then \(1/\lambda^*\) is also an eigenvalue. Eigenvalues on the unit circle are self-paired and correspond to propagating Floquet--Bloch branches. Their currents split equally into positive and negative signs. Eigenvalues away from the unit circle occur in reciprocal-conjugate pairs and correspond to right- and left-decaying evanescent branches.}
    \label{fig:symp_spec}
\end{figure}

The next structural input is the sign balance of propagating branches.

\begin{theorem}[Current-sign equal division]
\label{thm:tm_sign}
Fix a regular quasienergy at which no propagating branch has zero current.
Then the propagating Floquet--Bloch branches split into positive- and negative-current sectors with equal multiplicities:
\begin{equation}
N_+(\epsilon)=N_-(\epsilon)\equiv N_B(\epsilon).
\label{eq:tm_sign_balance}
\end{equation}
Equivalently, the right-going and left-going propagating channel spaces have the same dimension \cite{Mehrmann2016}.
\end{theorem}

A proof is given in Appendix~\ref{appsubsec:tm_sign}.
Combined with reciprocal-conjugate pairing for evanescent branches, this implies that the full local channel decomposition can be organized into two equally sized sectors, denoted \(\oplus\) and \(\ominus\): the \(\oplus\) sector contains all right-propagating and right-decaying modes, while the \(\ominus\) sector contains all left-propagating and left-decaying modes.
This is the channel grouping used later in the interface-scattering formalism.

\subsection{Pair-adapted dual vectors in the nondegenerate case}
\label{subsec:tm_pair_dual}

The scattering construction requires a dual basis adapted to the \(J\)-metric rather than to the Euclidean inner product.
In the nondegenerate case, this dual structure is most naturally expressed in terms of the reciprocal-conjugate pairing introduced above.

Assume first that all relevant eigenvalues are nondegenerate.
For each eigenvector \(y_\mu\), choose its paired vector \(\bar y_\mu\) according to Eqs.~\eqref{eq:tm_pair_def}--\eqref{eq:tm_pair_involution}.
Then Lemma~\ref{lem:tm_Jorth} implies
\[
\bar y_\mu^\dagger J y_\nu=0,
\qquad
\nu\neq\mu,
\]
because in the nondegenerate case the only eigenvalue satisfying
\[
\bar\lambda_\mu^*\lambda_\nu=1
\]
is \(\lambda_\nu=\lambda_\mu\).
Accordingly, one may define the pair-adapted dual row by
\begin{equation}
y_\mu^\ddagger
=
\frac{\bar y_\mu^\dagger J}{\bar y_\mu^\dagger J y_\mu}.
\label{eq:tm_pair_dual_def}
\end{equation}
By construction,
\begin{equation}
y_\mu^\ddagger y_\nu=\delta_{\mu\nu}.
\label{eq:tm_pair_dual_orth}
\end{equation}

For a self-paired propagating branch, \(\bar y_\mu=y_\mu\), so Eq.~\eqref{eq:tm_pair_dual_def} reduces to the familiar \(J\)-normalization of a single branch:
\begin{equation}
y_\mu^\ddagger
=
\frac{y_\mu^\dagger J}{y_\mu^\dagger J y_\mu}.
\label{eq:tm_selfpair_dual}
\end{equation}
For a non-self-paired evanescent branch, the dual row is built from its reciprocal-conjugate partner.

Thus, in the nondegenerate case, the operation \(\ddagger\) is simply:
take the reciprocal-conjugate partner, then normalize its \(J\)-pairing with the original vector.
This is the form used most often in the main text.
The only complication arises in degenerate reciprocal-conjugate classes, where one must first recombine the basis inside each class.

\subsection{\(J\)-biorthogonal basis and completeness}
\label{subsec:tm_dual_basis}

We now state the general completeness result in a form that also covers degenerate reciprocal-conjugate classes.

\begin{proposition}[\(J\)-biorthogonal basis and completeness]
\label{prop:tm_biorth}
Assume \(F\) is diagonalizable and satisfies \(F^\dagger JF=J\).
Then, after possible linear recombination inside degenerate reciprocal-conjugate eigenspaces, one can construct eigenvectors \(\{\tilde y_\alpha\}_{\alpha=1}^{2N}\) and dual rows \(\{\tilde y_\alpha^{\ddagger}\}_{\alpha=1}^{2N}\) such that
\begin{equation}
\tilde y_\alpha^{\ddagger}\tilde y_\beta=\delta_{\alpha\beta},
\qquad
\sum_{\alpha=1}^{2N}\tilde y_\alpha \tilde y_\alpha^{\ddagger}=I_{2N}.
\label{eq:tm_dual_complete}
\end{equation}
\end{proposition}

In the nondegenerate case, Proposition~\ref{prop:tm_biorth} reduces directly to the pair-adapted dual construction \eqref{eq:tm_pair_dual_def}.
The detailed blockwise construction for general degenerate reciprocal-conjugate classes is given in Appendix~\ref{appsubsec:tm_dual}.
This \(J\)-biorthogonal completeness is the algebraic basis for the closed-form matching formulas used later in Sec.~\ref{sec:interface_scattering}.
\subsection{Unit-cell probability, current, and group velocity}
\label{subsec:tm_nvj}

For a regular simple propagating branch, the transfer-matrix current is tied directly to the Floquet--Bloch group velocity.

\begin{proposition}[Unit-cell probability, current, and group velocity]
\label{prop:tm_nvj}
Let \(Y_\mu(x,\epsilon)=[\Phi_\mu(x,\epsilon),\Phi_\mu'(x,\epsilon)]^T\) be a regular simple propagating bulk branch with eigenvalue \(\lambda_\mu(\epsilon)=e^{ik_\mu(\epsilon)}\).
Define the time-averaged probability in one unit cell by
\begin{equation}
\begin{split}
\rho_\mu(\epsilon)
&= \int_0^1 \Phi_\mu(x,\epsilon)^\dagger \Phi_\mu(x,\epsilon)\,dx \\
&= \frac1\tau\int_0^\tau dt\int_0^1 dx\,|\psi_{\mu,\epsilon}(x,t)|^2.
\end{split}
\label{eq:tm_cell_prob}
\end{equation}
Then
\begin{equation}
\rho_\mu(\epsilon)\,v_\mu(\epsilon)=j_\mu(\epsilon),
\qquad
v_\mu(\epsilon)=\frac{\partial\epsilon}{\partial k_\mu},
\label{eq:tm_nvj}
\end{equation}
where \(j_\mu(\epsilon)=-i\,Y_\mu^\dagger JY_\mu\) is the conserved current.
\end{proposition}

The proof is given in Appendix~\ref{appsubsec:tm_nvj}.
Equation~\eqref{eq:tm_nvj} identifies the sign of the conserved current with the sign of the group velocity for regular simple propagating branches.
It is also the bridge between transfer-matrix current normalization and the spectral normalization of generalized states.

\subsection{Current normalization and spectral normalization}
\label{subsec:tm_current_norm}

We finally connect the transfer-matrix current normalization to the usual spectral normalization of extended states.

\begin{proposition}[Unit-current normalization from spectral normalization]
\label{prop:tm_current_en}
Fix a regular propagating Floquet--Bloch branch
\(\epsilon=\epsilon_\mu(k)\) with \(v_\mu(\epsilon)\neq0\).
Let \(u_{\mu k}(x,t)\) be normalized in one space-time unit cell,
\begin{equation}
\frac1\tau\int_0^\tau dt\int_0^1 dx\,
u_{\mu k}(x,t)^*u_{\nu k}(x,t)=\delta_{\mu\nu}.
\label{eq:tm_cell_norm}
\end{equation}

The double bracket denotes the extended Floquet inner product with time averaged over one driving period and space integrated over the full line:
\begin{equation}
\langle\!\langle \psi\mid \phi\rangle\!\rangle
=
\frac{1}{\tau}\int_0^\tau dt\int_{\mathbb R} dx\,
\psi(x,t)^*\phi(x,t),
\label{eq:tm_double_bracket}
\end{equation}
with the usual distributional interpretation for spatially extended generalized states. 
For cell-periodic functions, the corresponding cell inner product is obtained by restricting the spatial integral to one unit cell.

With the cell-periodic Floquet--Bloch functions normalized by Eq.~\eqref{eq:tm_cell_norm}, define the spatially extended \(k\)-normalized Bloch state
\begin{equation}
\Phi_{\mu k}(x,t)
=
\frac{1}{\sqrt{2\pi}} e^{ikx} u_{\mu k}(x,t).
\label{eq:tm_Phi_k_def}
\end{equation}
Then
\begin{equation}
\langle\!\langle \Phi_{\mu k}\mid \Phi_{\nu k'}\rangle\!\rangle
=
\delta_{\mu\nu}\delta(k-k').
\label{eq:tm_k_norm}
\end{equation}

Changing variables from \(k\) to quasienergy gives
\begin{equation}
\delta(k-k')
=
|v_\mu(\epsilon)|\,\delta(\epsilon-\epsilon'),
\qquad
v_\mu(\epsilon)=\frac{\partial\epsilon_\mu}{\partial k}.
\label{eq:tm_delta_k_delta_epsilon}
\end{equation}
Hence the quasienergy-normalized generalized state is
\begin{equation}
\Upsilon_\mu^{\en}(\epsilon)
=
\frac{1}{\sqrt{|v_\mu(\epsilon)|}}\,
\Phi_{\mu,k_\mu(\epsilon)},
\label{eq:tm_energy_norm}
\end{equation}
and satisfies
\begin{equation}
\langle\!\langle
\Upsilon_\mu^{\en}(\epsilon)
\mid
\Upsilon_\nu^{\en}(\epsilon')
\rangle\!\rangle
=
\delta_{\mu\nu}\delta(\epsilon-\epsilon').
\label{eq:tm_energy_delta_norm}
\end{equation}

The corresponding unit-current normalized state is
\begin{equation}
\Upsilon_\mu^{\cur}(\epsilon)
=
\sqrt{2\pi}\,\Upsilon_\mu^{\en}(\epsilon),
\label{eq:tm_current_norm}
\end{equation}
which has unit current magnitude,
\begin{equation}
|j_\mu(\epsilon)|=1.
\label{eq:tm_unit_current}
\end{equation}
Hence, a unit-current propagating bulk mode is exactly the same object as the quasienergy-normalized state up to a constant scaling.
\end{proposition}

The proof is given in Appendix~\ref{appsubsec:tm_currentnorm}.
This equivalence is what will later allow us to formulate the open sector in terms of physical unit-current scattering states while still computing everything in the transfer-matrix language.
\section{Interface scattering, finite-length composition, and smoothed transmittance}
\label{sec:interface_scattering}

In this section we assemble the scattering construction used in the rest of the paper.
The structural input from Sec.~\ref{sec:tm_csp} is threefold:
(i) each homogeneous region admits a channel basis adapted to the \(J\)-metric;
(ii) the full local channel decomposition is into \(\oplus\) and \(\ominus\) sectors, where \(\oplus\) contains all right-propagating and right-decaying modes and \(\ominus\) contains all left-propagating and left-decaying modes; the sign balance of propagating branches together with the reciprocal pairing of evanescent branches ensures that these sectors have equal dimension and that interface matching is solvable;
(iii) the \(J\)-biorthogonal dual basis provides closed algebraic formulas for interface matching.

\subsection{Channel decomposition in a homogeneous region}
\label{subsec:channel_decomp}

Fix a regular quasienergy \(\epsilon\).
For any homogeneous semi-infinite region \(X\), the one-cell transfer matrix \(F_X(\epsilon)\) is diagonalizable in the regular regime.
Its eigenvectors therefore provide a complete local channel basis, naturally organized by propagation/decay direction.
We write
\begin{equation}
X(\epsilon)=[\,X_\oplus(\epsilon)\mid X_\ominus(\epsilon)\,],
\label{eq:scatt_X_basis}
\end{equation}
where
\begin{itemize}
\item \(X_\oplus\) collects modes that are either propagating to the right or decaying as \(x\to +\infty\),
\item \(X_\ominus\) collects modes that are either propagating to the left or decaying as \(x\to -\infty\).
\end{itemize}
Equivalently, after diagonalizing \(F_X(\epsilon)\), one obtains a block decomposition into the \(\oplus\) and \(\ominus\) sectors together with the corresponding diagonal propagation matrices.
This is precisely the structure used below for interface matching and, later, for the finite-segment scattering problem.
\begin{figure}
    \centering
    \includegraphics[width=1\linewidth]{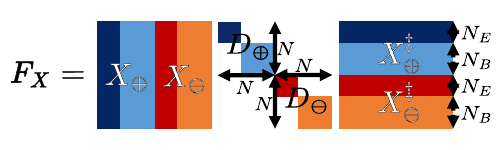}
    \caption{Schematic channel decomposition obtained from diagonalizing the one-cell transfer matrix in a homogeneous region. 
    The eigenvectors are organized into the \(\oplus\) and \(\ominus\) sectors according to propagation/decay direction, and the associated diagonal propagation matrices are obtained at the same time. 
    This is the basis structure used for interface matching and finite-segment scattering below.}
    \label{fig:diagonal}
\end{figure}
Let
\begin{equation}
X^{\ddagger}(\epsilon)=
\begin{bmatrix}
X_\oplus^{\ddagger}(\epsilon)\\
X_\ominus^{\ddagger}(\epsilon)
\end{bmatrix}
\label{eq:scatt_X_dual}
\end{equation}
denote the \(J\)-adapted dual basis constructed in Sec.~\ref{subsec:tm_dual_basis}.
Then
\begin{equation}
X^{\ddagger}X=I,
\qquad
X_\sigma^{\ddagger}X_{\sigma'}=\delta_{\sigma\sigma'}I,
\qquad
\sigma,\sigma'\in\{\oplus,\ominus\}.
\label{eq:scatt_X_duality}
\end{equation}

For propagating channels, the sign of the current distinguishes transport direction.
By Theorem~\ref{thm:tm_sign}, the right-going and left-going propagating sectors have equal dimensions.
This is the structural reason why the open scattering problem closes on square incoming and outgoing channel blocks.

\subsection{Interface scattering between two semi-infinite regions}
\label{subsec:interface_scattering}

Consider a sharp interface at \(x=0\) between a left semi-infinite region \(A\) and a right semi-infinite region \(B\).
Let
\[
A=[\,A_\oplus\mid A_\ominus\,],
\qquad
B=[\,B_\oplus\mid B_\ominus\,]
\]
be the corresponding full channel bases at the interface, where \(\oplus\) contains all right-propagating and right-decaying modes and \(\ominus\) contains all left-propagating and left-decaying modes.

\begin{figure}
    \centering
    \includegraphics[width=1\linewidth]{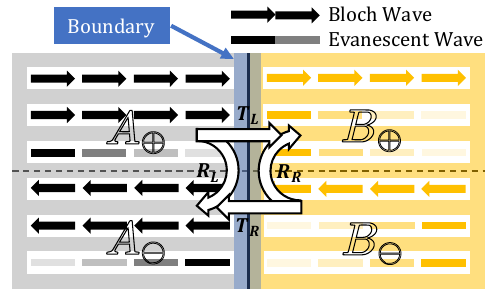}
    \caption{Scattering at an  interface between two semi-infinite homogeneous regions \(A\) and \(B\).}
    \label{fig:interface}
\end{figure}

For left incidence, the matching condition reads
\begin{equation}
A_\oplus\,a_{\mathrm{in}} + A_\ominus\,a_{\mathrm{ref}}
=
B_\oplus\,b_{\mathrm{tr}},
\label{eq:scatt_interface_left_match}
\end{equation}
where \(a_{\mathrm{in}}\) is the incoming amplitude in region \(A\), \(a_{\mathrm{ref}}\) is the reflected amplitude in region \(A\), and \(b_{\mathrm{tr}}\) is the transmitted amplitude in region \(B\).
Applying the dual rows of \(A\) gives
\begin{equation}
a_{\mathrm{in}}=A_\oplus^{\ddagger}B_\oplus\,b_{\mathrm{tr}},
\qquad
a_{\mathrm{ref}}=A_\ominus^{\ddagger}B_\oplus\,b_{\mathrm{tr}}.
\label{eq:scatt_interface_left_dual}
\end{equation}
Reciprocal-conjugate pairing in the evanescent sector and sign balance in the propagating sector ensure that the \(+\) sectors on the two sides have the same dimension, and likewise for the \(-\) sectors.
Accordingly, both \(A_\oplus^{\ddagger}B_\oplus\) and \(B_\ominus^{\ddagger}A_\ominus\) are square.
Whenever \(A_\oplus^{\ddagger}B_\oplus\) is invertible, the interface transmission and reflection matrices are
\begin{equation}
T_L=(A_\oplus^{\ddagger}B_\oplus)^{-1},
\qquad
R_L=A_\ominus^{\ddagger}B_\oplus\,T_L.
\label{eq:scatt_interface_left}
\end{equation}

For right incidence, the matching condition is
\begin{equation}
A_\ominus\,a_{\mathrm{tr}}
=
B_\ominus\,b_{\mathrm{in}} + B_\oplus\,b_{\mathrm{ref}},
\label{eq:scatt_interface_right_match}
\end{equation}
which gives
\begin{equation}
T_R=(B_\ominus^{\ddagger}A_\ominus)^{-1},
\qquad
R_R=B_\oplus^{\ddagger}A_\ominus\,T_R,
\label{eq:scatt_interface_right}
\end{equation}
provided \(B_\ominus^{\ddagger}A_\ominus\) is invertible.

Equations~\eqref{eq:scatt_interface_left} and \eqref{eq:scatt_interface_right} are the four basic scattering blocks of a sharp interface.

\medskip

We now extend this to a finite boundary region.
Suppose a boundary potential occupies a finite interval and interpolates between region \(A\) on the left and region \(B\) on the right.
If the boundary is sufficiently short, one may compute its scattering blocks directly from the transfer matrix across that interval by propagating between the two ends and projecting onto the channel bases on either side.
For example, if the boundary region occupies the interval \((x_1,x_2)\), then the left-incident transmission block is
\begin{equation}
T_L(\epsilon)
=
\bigl(
A_\oplus(x_1,\epsilon)^{\ddagger}
\,F(x_1,x_2;\epsilon)\,
B_\oplus(x_2,\epsilon)
\bigr)^{-1}.
\label{eq:scatt_interface_left_boundary}
\end{equation}
For long or strongly inhomogeneous boundary regions, however, this direct transfer-matrix route can be numerically unstable.
A more stable method is to compute the scattering matrices of shorter pieces and then concatenate them.
Suppose a first finite region \(1\) connects \(A\) to an intermediate homogeneous region \(B\), and a second finite region \(2\) connects \(B\) to a final homogeneous region \(C\).
Let
\[
(T_L^1,R_L^1,R_R^1,T_R^1)
\]
be the four scattering blocks of region \(1\), and
\[
(T_L^2,R_L^2,R_R^2,T_R^2)
\]
those of region \(2\).
Then the total scattering matrix from \(A\) to \(C\) is obtained by summing the infinite sequence of multiple reflections in the infinitesimal intermediate region \(B\):
\begin{equation}
T_L
=
T_L^2\bigl(I-R_R^1R_L^2\bigr)^{-1}T_L^1,
\label{eq:scatt_merge_TL}
\end{equation}
\begin{equation}
R_L
=
R_L^1
+
T_R^1R_L^2\bigl(I-R_R^1R_L^2\bigr)^{-1}T_L^1,
\label{eq:scatt_merge_RL}
\end{equation}
\begin{equation}
R_R
=
R_R^2
+
T_L^2R_R^1\bigl(I-R_L^2R_R^1\bigr)^{-1}T_R^2,
\label{eq:scatt_merge_RR}
\end{equation}
\begin{equation}
T_R
=
T_R^1\bigl(I-R_L^2R_R^1\bigr)^{-1}T_R^2.
\label{eq:scatt_merge_TR}
\end{equation}

\begin{figure}
    \centering
    \includegraphics[width=1\linewidth]{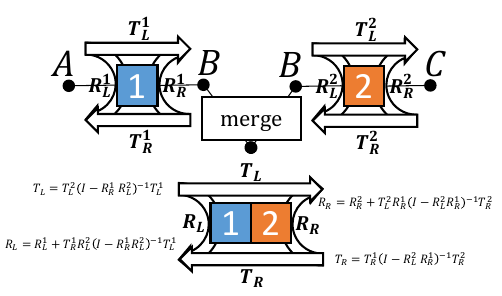}
    \caption{Concatenation of scattering matrices. Two finite regions are merged by summing the infinite sequence of reflections in the intermediate homogeneous region \(B\).}
    \label{fig:Smerge}
\end{figure}

They provide a numerically stable way to build the scattering matrix of an extended boundary from shorter pieces, and the same multiple-reflection logic will reappear in the finite-segment scattering problem.

\subsection{Finite Floquet segment between two leads}
\label{subsec:finite_segment}

We now consider the geometry relevant to the open transport problem: a finite Floquet segment of length \(L\) unit cells, sandwiched between two undriven leads.
Let the left lead, central Floquet region, and right lead be denoted by \(\mathsf L\), \(\mathsf C\), and \(\mathsf R\), respectively.

A crucial point is that the exact finite-sample matching problem is first formulated in the \emph{full} channel basis, not in the propagating sector alone.
For the central homogeneous Floquet region, we therefore start from
\begin{equation}
\mathsf C(\epsilon)
=
\bigl[\mathsf C_\oplus(\epsilon)\mid \mathsf C_\ominus(\epsilon)\bigr],
\label{eq:scatt_C_basis}
\end{equation}
where \(\mathsf C_\oplus\) contains all right-propagating and right-decaying modes, while \(\mathsf C_\ominus\) contains all left-propagating and left-decaying modes.
More explicitly,
\begin{equation}
\mathsf C_\oplus
=
\bigl[\mathsf C_\oplus^{\decay}\mid \mathsf C_\oplus^{\prop}\bigr],
\qquad
\mathsf C_\ominus
=
\bigl[\mathsf C_\ominus^{\decay}\mid \mathsf C_\ominus^{\prop}\bigr],
\label{eq:scatt_C_split}
\end{equation}
with
\begin{equation}
\mathsf C_\oplus^{\prop}
=
\bigl[Y_{\alpha_1}^{+},\dots,Y_{\alpha_{N_B}}^{+}\bigr],
\qquad
\mathsf C_\ominus^{\prop}
=
\bigl[Y_{\beta_1}^{-},\dots,Y_{\beta_{N_B}}^{-}\bigr],
\label{eq:scatt_C_prop}
\end{equation}
and
\begin{equation}
\mathsf C_\oplus^{\decay}
=
\bigl[Y_{\gamma_1}^{+},\dots,Y_{\gamma_{N_E}}^{+}\bigr],
\qquad
\mathsf C_\ominus^{\decay}
=
\bigl[Y_{\gamma_1}^{-},\dots,Y_{\gamma_{N_E}}^{-}\bigr].
\label{eq:scatt_C_decay}
\end{equation}

These channels arise from the diagonalization of the one-cell transfer matrix of the central region,
\begin{equation}
F_{\mathsf C}(\epsilon)
=
\mathsf C(\epsilon)
\begin{bmatrix}
D_\oplus(\epsilon) & 0\\
0 & D_\ominus(\epsilon)
\end{bmatrix}
\mathsf C(\epsilon)^\ddagger.
\label{eq:scatt_FC_diag}
\end{equation}
For each evanescent partner pair labeled by \(\gamma\), we write
\begin{equation}
\lambda_{\gamma,+}(\epsilon)=e^{-\kappa_\gamma(\epsilon)^*},
\quad
\lambda_{\gamma,-}(\epsilon)=e^{\kappa_\gamma(\epsilon)},
\quad
\Re \kappa_\gamma(\epsilon)>0,
\label{eq:scatt_lambda_kappa_pair}
\end{equation}
so that \(\lambda_{\gamma,-}=1/\lambda_{\gamma,+}^*\).
Accordingly,
\begin{equation}
\begin{split}
D_\oplus(\epsilon) = \diag\!\Bigl(
    &e^{-\kappa_{\gamma_1}(\epsilon)^*}, \dots, e^{-\kappa_{\gamma_{N_E}}(\epsilon)^*}, \\
    &e^{ik_{\alpha_1}(\epsilon)}, \dots, e^{ik_{\alpha_{N_B}}(\epsilon)}
\Bigr),
\end{split}
\label{eq:scatt_Dplus_full}
\end{equation}
\begin{equation}
\begin{split}
D_\ominus(\epsilon) = \diag\!\Bigl(
    &e^{\kappa_{\gamma_1}(\epsilon)}, \dots, e^{\kappa_{\gamma_{N_E}}(\epsilon)}, \\
    &e^{ik_{\beta_1}(\epsilon)}, \dots, e^{ik_{\beta_{N_B}}(\epsilon)}
\Bigr).
\end{split}
\label{eq:scatt_Dminus_full}
\end{equation}
Thus \(D_\oplus^L\) propagates the full \(+\) sector across \(L\) cells, while \(D_\ominus^{-L}\) propagates the full \(-\) sector back across the sample.

For the undriven leads, the channels are free-space sideband solutions.
Let \(\mathbf e_n\) denote the canonical basis vector in frequency space, and define
\begin{equation}
E_n(\epsilon)=\epsilon+n\omega.
\label{eq:scatt_En}
\end{equation}
For \(E_n(\epsilon)>0\), let
\begin{equation}
k_n(\epsilon)=\sqrt{E_n(\epsilon)},
\label{eq:scatt_kn}
\end{equation}
and define the propagating lead channels by
\begin{equation}
Y_n^{\rightarrow,\free}(\epsilon)
=
\frac{1}{\sqrt{2k_n}}
\begin{bmatrix}
\mathbf e_n\\
ik_n\,\mathbf e_n
\end{bmatrix},
\label{eq:scatt_free_prop_right}
\end{equation}
\begin{equation}
Y_n^{\leftarrow,\free}(\epsilon)
=
\frac{1}{\sqrt{2k_n}}
\begin{bmatrix}
\mathbf e_n\\
-ik_n\,\mathbf e_n
\end{bmatrix}.
\label{eq:scatt_free_prop_left}
\end{equation}
For \(E_n(\epsilon)<0\), let
\begin{equation}
\kappa_n(\epsilon)=\sqrt{-E_n(\epsilon)},
\label{eq:scatt_kappan}
\end{equation}
and define the evanescent lead channels by
\begin{equation}
Y_n^{+,\decay,\free}(\epsilon)
=
\frac{1}{\sqrt{2\kappa_n}}
\begin{bmatrix}
\mathbf e_n\\
-\kappa_n\,\mathbf e_n
\end{bmatrix},
\label{eq:scatt_free_decay_plus}
\end{equation}
\begin{equation}
Y_n^{-,\decay,\free}(\epsilon)
=
\frac{1}{\sqrt{2\kappa_n}}
\begin{bmatrix}
\mathbf e_n\\
\kappa_n\,\mathbf e_n
\end{bmatrix}.
\label{eq:scatt_free_decay_minus}
\end{equation}
Collecting these channels gives the free lead bases
\begin{equation}
\begin{split}
\mathsf L^{\free}(\epsilon) &= 
   \bigl[\mathsf L_\oplus^{\free}(\epsilon)\mid \mathsf L_\ominus^{\free}(\epsilon)\bigr], \\
\mathsf R^{\free}(\epsilon) &= 
   \bigl[\mathsf R_\oplus^{\free}(\epsilon)\mid \mathsf R_\ominus^{\free}(\epsilon)\bigr].
\end{split}
\label{eq:scatt_free_lead_basis}
\end{equation}

\begin{figure}
    \centering
    \includegraphics[width=1\linewidth]{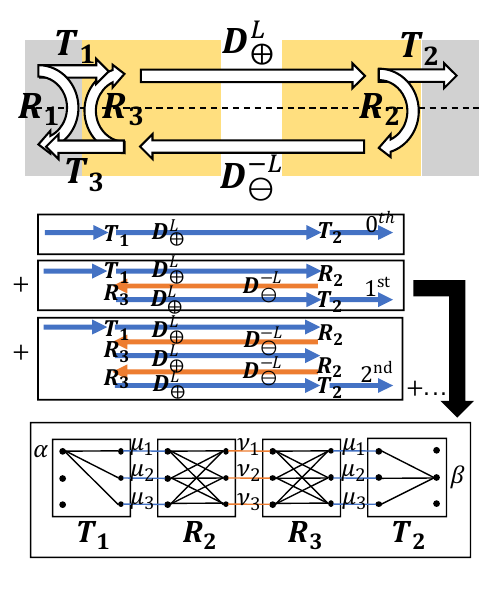}
    \caption{Scattering through a finite Floquet segment between two free leads. The figure illustrates both the loop-summation picture and its equivalent matrix-product representation. The same geometric-series structure will be used below in the long-sample smoothing analysis.}
    \label{fig:Triregion}
\end{figure}

Once the channel spaces of the left lead, the central Floquet region, and the right lead have been constructed, any finite boundary region between two such spaces can be treated as an interface-scattering problem in the corresponding full channel bases, with all frequency-channel couplings kept explicitly.
As discussed in Sec.~\ref{subsec:interface_scattering}, this may be done either directly from the transfer matrix of the boundary region or, more stably, by concatenating shorter pieces through scattering-matrix composition.
We therefore assume from now on that the effective scattering blocks of the left and right boundaries have already been obtained.
As an example, we only write the left-incident, right-outgoing transmission problem; the naming of all interface blocks is indicated in Fig.~\ref{fig:Triregion} and will not be repeated here.

With this convention, the exact left-to-right transmission matrix of the finite sample takes the multiple-reflection form
\begin{equation}
\begin{split}
T(\epsilon;L) &= T_2(\epsilon) \\
&\quad \times \Bigl(
    I - D_\oplus(\epsilon)^L R_3(\epsilon)
    D_\ominus(\epsilon)^{-L} R_2(\epsilon)
\Bigr)^{-1} \\
&\quad \times D_\oplus(\epsilon)^L T_1(\epsilon).
\end{split}
\label{eq:scatt_three_segment_full}
\end{equation}
Equivalently, in the domain where the geometric series converges,
\begin{equation}
T(\epsilon;L)
=
T_2
\sum_{n=0}^{\infty}
\bigl(
D_\oplus^L
R_3
D_\ominus^{-L}
R_2
\bigr)^n
D_\oplus^L
T_1.
\label{eq:scatt_three_segment_series_full}
\end{equation}
This is the exact full-channel transmission formula.
Its physical content is the loop expansion shown in Fig.~\ref{fig:Triregion}: after entering the central region, the wave may undergo arbitrarily many round trips between the two boundaries before finally exiting into the right lead.

For the long-sample asymptotics relevant below, the central evanescent channels only contribute near the two boundaries.
Any scattering path that crosses the sample through a central evanescent component carries a factor \(e^{-\Re \kappa_\gamma(\epsilon)L}\), and is therefore exponentially small in \(L\).
Thus the exact finite-sample problem is first formulated in the full channel space, while in the long-sample limit one may simplify the central-region problem by cutting it to the propagating sector.

Writing
\begin{equation}
D_\oplus^{\prop}(\epsilon)
=
\diag\!\Bigl(
e^{ik_{\alpha_1}(\epsilon)},
\dots,
e^{ik_{\alpha_{N_B}}(\epsilon)}
\Bigr),
\label{eq:scatt_Dplus_prop}
\end{equation}
\begin{equation}
D_\ominus^{\prop}(\epsilon)
=
\diag\!\Bigl(
e^{ik_{\beta_1}(\epsilon)},
\dots,
e^{ik_{\beta_{N_B}}(\epsilon)}
\Bigr),
\label{eq:scatt_Dminus_prop}
\end{equation}
and denoting the corresponding propagating-sector interface blocks by
\[
T_1^{\cut},\qquad T_2^{\cut},\qquad R_2^{\cut},\qquad R_3^{\cut},
\]
one obtains the cut transmission formula
\begin{equation}
\begin{split}
& T^{\cut}(\epsilon;L) = T_2^{\cut}(\epsilon) \\
& \qquad \times \bigl(
    I - D_\oplus^{\prop}(\epsilon)^L R_3^{\cut}(\epsilon)
    D_\ominus^{\prop}(\epsilon)^{-L} R_2^{\cut}(\epsilon)
\bigr)^{-1} \\
& \qquad \times D_\oplus^{\prop}(\epsilon)^L T_1^{\cut}(\epsilon).
\end{split}
\label{eq:scatt_three_segment}
\end{equation}
This cut formula is the natural starting point for the long-sample analysis.
It keeps the propagating loops responsible for the nontrivial interference structure.

\subsection{Smoothed transmittance in the long-sample limit}
\label{subsec:smoothed_trans}

For finite \(L\), the cut transmission formula \eqref{eq:scatt_three_segment} exhibits rapid Fabry--P\'erot oscillations as functions of quasienergy and sample length \cite{Born1999}.
As \(L\) increases, these oscillations become increasingly dense, while their overall support and envelope remain tied to the same underlying branch-connectivity structure.
Accordingly, the physically relevant transport observable is not the pointwise finite-\(L\) line shape, but a local quasienergy average taken at fixed \(L\), followed by the long-sample limit \(L\to\infty\).
This same logic will be used throughout the paper: first one forms a shrinking quasienergy average, and only then takes the long-sample limit.

\begin{figure}
    \centering
    \includegraphics[width=1\linewidth]{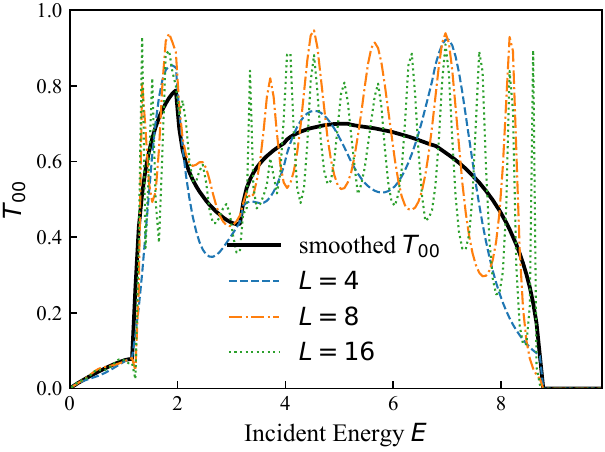}
    \caption{Illustration of shrinking-window smoothing for a fixed channel-resolved transmittance. Panels (a)--(c) show the pointwise finite-sample transmittance \(\mathcal T_{\beta\alpha}(\epsilon;L)\) for several lengths \(L\) (for example \(L=2,4,8\)). As \(L\) increases, the Fabry--P\'erot oscillations become progressively denser, but they fluctuate around a stable underlying profile. Panel (d) shows the corresponding smoothed long-sample transmittance \(\bar{\mathcal T}_{\beta\alpha}(\epsilon)\), which captures precisely this stable profile.}
    \label{fig:smoothed_trans_demo}
\end{figure}

Figure~\ref{fig:smoothed_trans_demo} gives a direct numerical illustration of this point.
The finite-\(L\) transmission curves become more oscillatory with increasing length, yet they do not wander arbitrarily:
they remain organized around a well-defined smooth profile.
The smoothed transmittance is precisely the systematic version of this profile.
In this sense, the shrinking-window procedure does not discard physical information; it removes the nonuniversal interference fringes while retaining the local spectral transport weight.

For propagating incoming and outgoing lead channels \(\alpha\) and \(\beta\), define the finite-sample channel-resolved transmittance
\begin{equation}
\mathcal T_{\beta\alpha}(\epsilon;L)
=
|T_{\beta\alpha}(\epsilon;L)|^2\,
\frac{|j_\beta(\epsilon)|}{|j_\alpha(\epsilon)|}.
\label{eq:scatt_finite_Tba}
\end{equation}
In the left-to-right transport setting considered later, both currents are positive and the absolute values may be dropped.
We keep them here so that the definition is independent of the directional labeling convention.

We now introduce the shrinking-window average in a form that will also be used later for branch-resolved bulk quantities.
For each \(L\), let \(W_L(\epsilon;\epsilon_0)\) be a normalized energy-envelope amplitude centered at \(\epsilon_0\), satisfying
\begin{equation}
\int d\epsilon\, |W_L(\epsilon;\epsilon_0)|^2 = 1,
\label{eq:scatt_W_norm}
\end{equation}
and supported, or effectively localized, in a quasienergy window of width \(\Delta\epsilon(L)\) such that
\begin{equation}
\Delta\epsilon(L)\to0,
\qquad
L\,\Delta\epsilon(L)\to\infty,
\qquad
L\to\infty.
\label{eq:scatt_window_scaling}
\end{equation}
The first condition preserves locality in quasienergy, while the second ensures that phases whose energy derivatives are of order \(L\) are still washed out by the averaging.
To prevent the envelope itself from introducing rapid variations that could compensate the fast oscillations, we further require
\begin{equation}
\partial_\epsilon\bigl(|W_L|^2\bigr) < C\,\Delta\epsilon^{-2},
\label{eq:slow_oscillation_W}
\end{equation}
which guarantees that \(|W_L|^2\) varies slowly and carries no Fourier components of order \(L\).

In the path representation used below, all \(O(L)\) quasienergy oscillations are carried explicitly by the propagation matrices \(D_\oplus^L\) and \(D_\ominus^{-L}\), while the remaining prefactors vary only on \(O(1)\) quasienergy scales across the shrinking window.
The final smoothed observables are representation independent; this assumption is used only to make the fast/slow separation explicit.

We therefore define the averaged transmittance at fixed \(L\) by
\begin{equation}
\avg{\mathcal T_{\beta\alpha}}^{(L)}(\epsilon_0)
=
\int d\epsilon\,
|W_L(\epsilon;\epsilon_0)|^2\,
\mathcal T_{\beta\alpha}(\epsilon;L),
\label{eq:scatt_avg_Tba}
\end{equation}
and the corresponding long-sample smoothed transmittance by
\begin{equation}
\bar{\mathcal T}_{\beta\alpha}(\epsilon_0)
=
\lim_{L\to\infty}
\avg{\mathcal T_{\beta\alpha}}^{(L)}(\epsilon_0),
\label{eq:scatt_smoothed_trans}
\end{equation}
provided the limit exists.
Although the finite-\(L\) average depends on the chosen admissible envelope \(W_L\), the long-sample limit will be shown below to depend only on the center \(\epsilon_0\), not on the detailed admissible shape of the envelope.

The relevance of the scaling \eqref{eq:scatt_window_scaling} is that the multiple-reflection expansion of Fig.~\ref{fig:Triregion} naturally decomposes \(T_{\beta\alpha}(\epsilon;L)\) into a sum over scattering paths, each carrying a phase whose quasienergy derivative is a total Wigner delay of order \(L\).
After the averaging \eqref{eq:scatt_avg_Tba}, only interference terms within a common delay sector survive.

\begin{proposition}[Delay-sector form of the smoothed transmittance]
\label{prop:delay_sector}
Assume the path expansion described in Appendix~\ref{appsubsec:smoothing_path}, in an admissible representation on a regular quasienergy window around \(\epsilon_0\).
Then
\begin{equation}
\bar{\mathcal T}_{\beta\alpha}(\epsilon_0)
=
\lim_{L\to\infty}
\frac{|j_\beta(\epsilon_0)|}{|j_\alpha(\epsilon_0)|}
\sum_{\tau}
\left|
\sum_{s\in\mathcal S_\tau}
A_{\beta\alpha}^{s}(\epsilon_0)\,
e^{i\phi_{\beta\alpha}^{s}(\epsilon_0;L)}
\right|^2,
\label{eq:scatt_delay_sector}
\end{equation}
where \(\mathcal S_\tau\) is the set of scattering paths with total Wigner delay \(\tau\) \cite{Wigner1955,Smith1960,Texier2016}.
In particular, the long-sample limit is independent of the detailed admissible choice of the energy-envelope amplitude \(W_L\).

Under the additional generic condition that equal delay implies equal path-count vector \(\mathbf N\), Eq.~\eqref{eq:scatt_delay_sector} may be rewritten as
\begin{equation}
\bar{\mathcal T}_{\beta\alpha}(\epsilon_0)
=
\frac{|j_\beta(\epsilon_0)|}{|j_\alpha(\epsilon_0)|}
\sum_{\mathbf N}
\left|
\sum_{s\in\mathcal S_{\mathbf N}}
A_{\beta\alpha}^{s}(\epsilon_0)
\right|^2,
\label{eq:scatt_counting_vector}
\end{equation}
where \(\mathbf N\) counts the numbers of traversals of the propagating branches.
\end{proposition}

The proof is given in Appendix~\ref{app:smoothing}.
The same shrinking-window logic will reappear later in the open-sector theory.
There, the endpoint of the path is no longer an outgoing lead channel \(\beta\), but a propagating bulk branch \(\mu\) in the central Floquet region.
One then works with branch-resolved amplitudes \(d_{\mu\alpha}(\epsilon;L)\) and pointwise intensities
\begin{equation}
\mathcal I_{\mu\alpha}(\epsilon;L)=|d_{\mu\alpha}(\epsilon;L)|^2.
\label{eq:scatt_I_mualpha_preview}
\end{equation}
As before, however, these pointwise quantities are not yet the relevant long-sample observables.
After shrinking-window averaging, they give the branch-resolved weights \(p_{\mu\alpha}(\epsilon_0)\), and after summing over incoming channels, \(p_\mu(\epsilon_0)\).
Thus the same smoothing mechanism that defines \(\bar{\mathcal T}_{\beta\alpha}\) here will later define the physically relevant open-sector quantities.

\section{Open sector as branch accessibility and escape}
\label{sec:open_sector}

We now turn from lead-to-lead transmission to the bulk quantity that controls robust open Floquet transport.
The relevant question is not only which asymptotic lead channels are open, but also which propagating Floquet--Bloch branches are actually accessed deep inside a long sample by physical scattering states generated from those channels.
The open-sector construction is therefore branch resolved from the outset.

For each physical open lead channel \(\alpha\), let
\begin{equation}
\Phi_{\alpha}^{\rm in,\en}(\epsilon),
\qquad
\Phi_{\beta}^{\rm out,\en}(\epsilon)
\label{eq:open_free_channel_states}
\end{equation}
denote the quasienergy-normalized free incoming and outgoing lead states, with
\begin{equation}
\langle\!\langle \Phi_{\alpha}^{\rm in,\en}(\epsilon)\mid
\Phi_{\alpha'}^{\rm in,\en}(\epsilon')\rangle\!\rangle
=
\delta_{\alpha\alpha'}\delta(\epsilon-\epsilon'),
\label{eq:open_free_channel_norm_in}
\end{equation}
and similarly
\begin{equation}
\langle\!\langle \Phi_{\beta}^{\rm out,\en}(\epsilon)\mid
\Phi_{\beta'}^{\rm out,\en}(\epsilon')\rangle\!\rangle
=
\delta_{\beta\beta'}\delta(\epsilon-\epsilon').
\label{eq:open_free_channel_norm_out}
\end{equation}

Let wave operators \(\Omega_{-,L}\) and \(\Omega_{+,L}\) denote the incoming and outgoing scattering maps in the fixed-quasienergy representation.
They generate the corresponding physical scattering states from free states \cite{Newton2013,Kato1966}.
\begin{equation}
\begin{split}
\Psi_{\alpha}^{(-),\en}(\epsilon;L) &= \Omega_{-,L}\Phi_{\alpha}^{\rm in,\en}(\epsilon), \\
\Psi_{\beta}^{(+),\en}(\epsilon;L) &= \Omega_{+,L}\Phi_{\beta}^{\rm out,\en}(\epsilon).
\end{split}
\label{eq:open_scattering_states_pm}
\end{equation}

In the regular quasienergy window considered here, we adopt the standard Floquet scattering picture that the physical scattering states generated from the open lead channels exhaust the relevant absolutely continuous scattering sector.
Equivalently, the incoming and outgoing scattering constructions have the same range\cite{Howland1979,Howland2012},
\begin{equation}
\Ran(\Omega_{-,L})=\Ran(\Omega_{+,L}).
\label{eq:open_same_range}
\end{equation}
Thus they provide two descriptions of the same physical scattering sector of the open device.

In particular, each of the two scattering families separately inherits the orthogonality of its free asymptotic reference states: the states \(\Psi_{\alpha}^{(-),\en}\) inherit the orthogonality of the incoming free channels, while the states \(\Psi_{\beta}^{(+),\en}\) inherit that of the outgoing free channels.
The role of Eq.~\eqref{eq:open_same_range} is then to allow the same physical scattering sector to be described either in the incoming representation or in the outgoing one.

The set of physical open incoming channels at quasienergy \(\epsilon\) is denoted by \(\mathcal A_{\open}(\epsilon)\).

\subsection{Deep-bulk branch amplitudes}
\label{subsec:open_deep_bulk}

Fix a regular quasienergy window
\begin{equation}
\mathcal E_*=(\epsilon_0-\delta,\epsilon_0+\delta),
\label{eq:open_Estar}
\end{equation}
on which the number of propagating bulk branches is constant, all relevant group velocities stay nonzero, and the set of physical open lead channels does not change.
On this window, let
\begin{equation}
\bigl\{\Upsilon_\mu^{\cur}(x,t;\epsilon)\bigr\}
\label{eq:open_bulk_frame}
\end{equation}
be the current-normalized propagating Floquet--Bloch branches of the central homogeneous region.

For a long sample, choose a point \(x_c(L)\) in the central region such that
\begin{equation}
x_c(L), L-x_c(L)\to\infty
\qquad
\text{linearly with }L.
\label{eq:open_xc_bulk}
\end{equation}
In a neighborhood of \(x_c(L)\), every physical incoming scattering state admits a deep-bulk propagating decomposition
\begin{equation}
\begin{split}
\Psi_{\alpha}^{(-),\cur}(x,t;\epsilon;L)
&= \sum_{\mu} d_{\mu\alpha}(\epsilon;L)\, \Upsilon_\mu^{\cur}(x,t;\epsilon) \\
&\quad + \eta_\alpha(x,t;\epsilon;L).
\end{split}
\label{eq:open_deep_bulk_expansion}
\end{equation}
where \(\eta_\alpha\) denotes the evanescent component, which is exponentially small at the deep-bulk point in the long-sample limit.
Since the current-normalized and quasienergy-normalized branch states differ only by the constant factor established in Proposition~\ref{prop:tm_current_en}, the same expansion may equivalently be written in the quasienergy normalization as
\begin{equation}
\begin{split}
\Psi_{\alpha}^{(-),\en}(x,t;\epsilon;L)
&= \sum_{\mu} d_{\mu\alpha}(\epsilon;L)\, \Upsilon_\mu^{\en}(x,t;\epsilon) \\
&\quad + \frac{1}{\sqrt{2\pi}}\eta_\alpha(x,t;\epsilon;L).
\end{split}
\label{eq:open_deep_bulk_expansion_en}
\end{equation}

The coefficient \(d_{\mu\alpha}(\epsilon;L)\) is the quantity of direct physical interest:
it is the amplitude with which the physical incoming channel \(\alpha\) accesses branch \(\mu\) deep inside the sample.
Operationally, \(d_{\mu\alpha}\) is read off directly from the scattering solution in the central homogeneous region.
On a regular quasienergy window, distinct propagating branches carry distinct Bloch phases, so this decomposition is asymptotically unambiguous deep in the bulk.

\subsection{Branch-resolved intensities and shrinking-window limits}
\label{subsec:open_intensity}

The basic pointwise branch-resolved quantity is
\begin{equation}
\mathcal I_{\mu\alpha}(\epsilon;L)
=
|d_{\mu\alpha}(\epsilon;L)|^2.
\label{eq:open_Imualpha}
\end{equation}
Summing over all physical open incoming channels gives
\begin{equation}
\mathcal I_\mu(\epsilon;L)
=
\sum_{\alpha\in\mathcal A_{\open}(\epsilon)}
\mathcal I_{\mu\alpha}(\epsilon;L).
\label{eq:open_Imu}
\end{equation}

At fixed \((\epsilon,L)\), these quantities are not yet the physically relevant long-sample observables.
As in Sec.~\ref{subsec:smoothed_trans}, the relevant branch weights are obtained only after shrinking-window averaging.
We therefore use the same admissible energy-envelope amplitude \(W(\epsilon;\epsilon_0)\) as before, normalized by
\begin{equation}
\int d\epsilon\,|W(\epsilon;\epsilon_0)|^2=1,
\label{eq:open_W_norm}
\end{equation}
and localized in a window of width \(\Delta\epsilon(L)\) such that
\begin{equation}
\Delta\epsilon(L)\to0,
\qquad
L\,\Delta\epsilon(L)\to\infty,
\qquad
L\to\infty.
\label{eq:open_average_window}
\end{equation}
We then define
\begin{equation}
\avg{\mathcal I_{\mu\alpha}}^{(L)}(\epsilon_0)
=
\int d\epsilon\,
|W(\epsilon;\epsilon_0)|^2\,
\mathcal I_{\mu\alpha}(\epsilon;L),
\label{eq:open_avg_Imualpha}
\end{equation}
and
\begin{equation}
\avg{\mathcal I_{\mu}}^{(L)}(\epsilon_0)
=
\int d\epsilon\,
|W(\epsilon;\epsilon_0)|^2\,
\mathcal I_{\mu}(\epsilon;L).
\label{eq:open_avg_Imu}
\end{equation}

Their long-sample limits are
\begin{equation}
p_{\mu\alpha}(\epsilon_0)
=
\lim_{L\to\infty}\avg{\mathcal I_{\mu\alpha}}^{(L)}(\epsilon_0),
\label{eq:open_pmualpha_def}
\end{equation}
and
\begin{equation}
p_{\mu}(\epsilon_0)
=
\lim_{L\to\infty}\avg{\mathcal I_{\mu}}^{(L)}(\epsilon_0)
=
\sum_{\alpha\in\mathcal A_{\open}(\epsilon_0)} p_{\mu\alpha}(\epsilon_0),
\label{eq:open_pmu_def}
\end{equation}
whenever the limits exist.
At this stage, \(p_{\mu\alpha}\) and \(p_\mu\) are simply long-sample branch weights.

\subsection{Single-branch deep-bulk packets}
\label{subsec:open_packets}

Using the same admissible energy-envelope amplitude \(W(\epsilon;\epsilon_0)\), we now construct a wave packet on a single propagating branch \(\mu\), centered at the deep-bulk point \(x_c(L)\).
Since all previous shrinking-window observables depend on the envelope only through \(|W|^2\), we are free here to choose the phase of \(W\) so that the resulting packet is centered at the prescribed point \(x_c(L)\).

The corresponding single-branch deep-bulk packet is
\begin{equation}
\Psi_{\mu,W}^{(L)}(x,t)
=
\int d\epsilon\,
W(\epsilon;\epsilon_0)\,
\Upsilon_\mu^{\en}(x,t;\epsilon).
\label{eq:open_single_branch_packet}
\end{equation}
On a regular quasienergy window, the shrinking-window conditions
\[
\Delta\epsilon(L)\to0,
\qquad
L\,\Delta\epsilon(L)\to\infty
\]
imply, through the local dispersion relation, a spatial width \(\ell(L)\) satisfying
\begin{equation}
\ell(L)\to\infty,
\qquad
\ell(L)/L\to0.
\label{eq:open_packet_scale_limit}
\end{equation}
Thus the packet becomes broad on microscopic scales but remains asymptotically negligible compared with the sample length.
In particular, asymptotically all of its norm is concentrated in the deep bulk.

Because the packet is centered at a point whose distance to both boundaries grows linearly with \(L\), it probes only the deep-bulk part of the physical scattering states.
On the part of the packet carrying asymptotically all of its norm, every physical incoming scattering state is already given by the propagating decomposition \eqref{eq:open_deep_bulk_expansion}, up to an exponentially small remainder.
The evanescent boundary-local pieces have decayed exponentially by the time the packet reaches the deep bulk.
Thus, in the region actually sampled by the packet, the physical scattering states have converged to their \(d_{\mu\alpha}\)-weighted propagating Floquet--Bloch form.

In the present open geometry, any component that escapes the sample must ultimately appear in the free outgoing lead sector.
Within the scattering picture adopted above, the incoming and outgoing wave operators describe the same physical scattering sector.
The orthogonal projection onto that sector is therefore
\begin{equation}
P_{{\rm sc},L}
=
\Omega_{-,L}\Omega_{-,L}^\dagger
=
\Omega_{+,L}\Omega_{+,L}^\dagger.
\label{eq:open_Psc}
\end{equation}
For the packet \(\Psi_{\mu,W}^{(L)}\), its total weight in the scattering sector may thus be computed either from the incoming representation or from the outgoing one.

We denote this scattering-sector weight by
\begin{equation}
\mathcal P_\mu^{\rm sc}(\epsilon_0)
=
\lim_{L\to\infty}
\bigl\langle\!\bigl\langle
\Psi_{\mu,W}^{(L)}
\big|
P_{{\rm sc},L}
\big|
\Psi_{\mu,W}^{(L)}
\bigr\rangle\!\bigr\rangle.
\label{eq:open_branch_scattering_weight}
\end{equation}
Below we show that this quantity is equal to \(p_\mu(\epsilon_0)\) when evaluated in the incoming representation, and equal to the escape probability when evaluated in the outgoing representation.

\subsection{Main theorem: averaged branch weight equals escape probability}
\label{subsec:open_main}

We may now state the central branch-resolved result.

\begin{theorem}[Averaged branch weight equals escape probability]
\label{thm:open_branch_escape}
Let \(W(\epsilon;\epsilon_0)\) be an admissible energy envelope supported in a sufficiently small regular window \(\mathcal E_*\), so that the branch-resolved deep-bulk decomposition \eqref{eq:open_deep_bulk_expansion} is well defined there.
Suppose that the incoming and outgoing wave operators exist on the open-channel sector and satisfy
\begin{equation}
\Ran(\Omega_{-,L})=\Ran(\Omega_{+,L}).
\label{eq:open_same_range2}
\end{equation}
Then
\begin{equation}
\mathcal P_\mu^{\esc}(\epsilon_0)=p_\mu(\epsilon_0).
\label{eq:open_branch_escape_main}
\end{equation}
\end{theorem}
\begin{proof}
Consider the packet \(\Psi_{\mu,W}^{(L)}\) defined in \eqref{eq:open_single_branch_packet}.
Its total weight in the physical scattering sector is the common quantity \(\mathcal P_\mu^{\rm sc}(\epsilon_0)\) defined in \eqref{eq:open_branch_scattering_weight}.

We first evaluate \(\mathcal P_\mu^{\rm sc}\) in the incoming representation.
Using
\[
P_{{\rm sc},L}=\Omega_{-,L}\Omega_{-,L}^\dagger
\]
and inserting the resolution of identity on the incoming free open-channel reference space, we obtain
\begin{equation}
\begin{split}
& \bigl\langle\!\bigl\langle
\Psi_{\mu,W}^{(L)}
\big|
P_{{\rm sc},L}
\big|
\Psi_{\mu,W}^{(L)}
\bigr\rangle\!\bigr\rangle
= \\
& \quad \sum_{\alpha\in\mathcal A_{\open}(\epsilon)}
   \int d\epsilon\,
   \bigl|
   \langle\!\langle
   \Psi_\alpha^{(-),\en}(\epsilon;L)
   \mid
   \Psi_{\mu,W}^{(L)}
   \rangle\!\rangle
   \bigr|^2 .
\end{split}
\label{eq:open_proj_incoming_short}
\end{equation}

Now the packet is localized in the deep bulk, while the evanescent boundary-local pieces of the incoming scattering states are exponentially small there.
Hence, in the shrinking-window / long-sample limit, the overlap in \eqref{eq:open_proj_incoming_short} is determined only by the deep-bulk propagating part
\[
\Psi_{\alpha}^{(-),\en}(x,t;\epsilon;L)
\sim
\sum_{\nu}
d_{\nu\alpha}(\epsilon;L)\,
\Upsilon_\nu^{\en}(x,t;\epsilon).
\]
Because the packet is built on the single branch \(\mu\), and distinct propagating Bloch branches are orthogonal on the regular window except for nongeneric accidental degeneracies, only the \(\nu=\mu\) contribution survives.
Therefore
\begin{equation}
\begin{split}
& \bigl\langle\!\bigl\langle
\Psi_{\mu,W}^{(L)}
\big|
P_{{\rm sc},L}
\big|
\Psi_{\mu,W}^{(L)}
\bigr\rangle\!\bigr\rangle
\longrightarrow \\
& \quad \sum_{\alpha\in\mathcal A_{\open}(\epsilon)}
   \int d\epsilon\,
   |W(\epsilon;\epsilon_0)|^2
   |d_{\mu\alpha}(\epsilon;L)|^2 .
\end{split}
\label{eq:open_proj_to_I_direct}
\end{equation}
Using the definition of \(\mathcal I_\mu\), this becomes
\begin{equation}
\bigl\langle\!\bigl\langle
\Psi_{\mu,W}^{(L)}
\big|
P_{{\rm sc},L}
\big|
\Psi_{\mu,W}^{(L)}
\bigr\rangle\!\bigr\rangle
\longrightarrow
\int d\epsilon\,
|W(\epsilon;\epsilon_0)|^2
\mathcal I_\mu(\epsilon;L).
\label{eq:open_proj_to_Imu_direct}
\end{equation}
Taking \(L\to\infty\) gives
\begin{equation}
\mathcal P_\mu^{\rm sc}(\epsilon_0)=p_\mu(\epsilon_0).
\label{eq:open_scattering_weight_equals_pmu}
\end{equation}

We next evaluate the same quantity in the outgoing representation.
Using
\[
P_{{\rm sc},L}=\Omega_{+,L}\Omega_{+,L}^\dagger,
\]
the packet weight in the scattering sector is its total weight in the outgoing scattering family.
In the present open geometry, any component of the packet that escapes must asymptotically emerge as a free outgoing lead wave.
Therefore
\begin{equation}
\mathcal P_\mu^{\rm sc}(\epsilon_0)=\mathcal P_\mu^{\esc}(\epsilon_0).
\label{eq:open_scattering_weight_equals_escape}
\end{equation}
Combining \eqref{eq:open_scattering_weight_equals_pmu} and \eqref{eq:open_scattering_weight_equals_escape} yields
\[
\mathcal P_\mu^{\esc}(\epsilon_0)=p_\mu(\epsilon_0),
\]
which is Eq.~\eqref{eq:open_branch_escape_main}.
\end{proof}

For a normalized packet, the complementary probability is the weight remaining outside the scattering sector. Thus
\begin{equation}
\mathcal P_\mu^{\esc}(\epsilon_0)
=
1-\mathcal P_\mu^{\rm b}(\epsilon_0),
\label{eq:open_escape_bound_complement}
\end{equation}
where \(\mathcal P_\mu^{\rm b}\) denotes the probability carried by the bound sector. This complementary form is the natural starting point for the generic-openness discussion in the next section.

\begin{figure}
    \centering
    \includegraphics[width=1\linewidth]{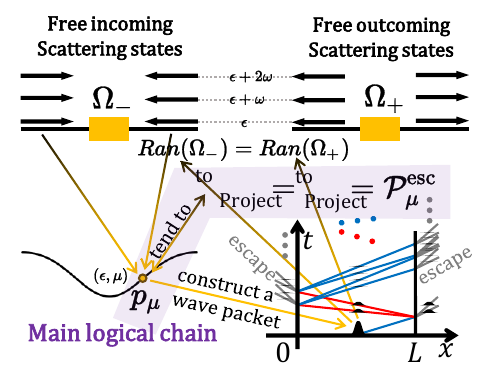}
    \caption{Logical structure of the open-sector construction. The incoming representation defines the branch-resolved long-sample weights \(p_{\mu\alpha}\) and the total branch weights \(p_\mu\), which quantify how physical incoming scattering states access propagating Floquet--Bloch branches deep inside a long sample. The outgoing representation yields the corresponding escape probability. The main result of this section is the identification \(p_\mu=\mathcal P_\mu^{\esc}\).}
\end{figure}

\section{Generic openness of propagating branches}
\label{sec:generic_openness}

By Theorem~\ref{thm:open_branch_escape}, the long-sample branch weight equals the escape probability,
\begin{equation}
p_\mu(\epsilon_0)=\mathcal P_\mu^{\esc}(\epsilon_0).
\label{eq:generic_p_equals_escape}
\end{equation}
For a normalized single-branch deep-bulk packet,
\begin{equation}
\mathcal P_\mu^{\esc}(\epsilon_0)
=
1-\mathcal P_\mu^{\rm b}(\epsilon_0),
\label{eq:generic_escape_bound}
\end{equation}
where \(\mathcal P_\mu^{\rm b}\) is the weight captured by the bound sector.
Hence
\begin{equation}
p_\mu(\epsilon_0)=1-\mathcal P_\mu^{\rm b}(\epsilon_0).
\end{equation}
For generic parameter values, true bound trapping of a propagating branch is absent, so that
\begin{equation}
\mathcal P_\mu^{\rm b}(\epsilon_0)=0
\qquad\text{and hence}\qquad
p_\mu(\epsilon_0)=1.
\end{equation}

We now explain why this is the generic situation for propagating branches.
At fixed regular quasienergy \(\epsilon\), a genuine Floquet bound state must decay in both leads.
Let \(B_L^-(\epsilon)\) be the matrix whose \(N_E^{\free}\) columns span the left-decaying free subspace.
On the right, let \(B_{R,\mathrm{forb}}(\epsilon)\) denote the forbidden sector, namely the direct sum of the wrong decaying channels and both propagating directions.
Its dimension is
\begin{equation}
N_E^{\free}+2N_B^{\free}.
\end{equation}
If \(B_{R,\mathrm{forb}}^\ddagger(\epsilon)\) is the corresponding dual row matrix, then a nonzero coefficient vector \(c\in\mathbb C^{N_E^{\free}}\) produces a genuine bound state only if the propagated right-end wave-function vector
\[
F(L,0;\epsilon)\,B_L^-(\epsilon)c
\]
has no forbidden component, i.e.
\begin{equation}
C(\epsilon)c=0,
\qquad
C(\epsilon):=
B_{R,\mathrm{forb}}^\ddagger(\epsilon)\,
F(L,0;\epsilon)\,
B_L^-(\epsilon).
\label{eq:generic_Cc0}
\end{equation}

The crucial point is that this differs qualitatively from an ordinary bound-state problem below the continuum.
When \(N_B^{\free}=0\), the matching matrix is square, and tuning the spectral parameter \(\epsilon\) can generically produce isolated bound-state energies, just as in the usual static problem.
By contrast, in an open quasienergy sector one has \(N_B^{\free}>0\), so \(C(\epsilon)\) is a tall rectangular matrix of size
\begin{equation}
\bigl(N_E^{\free}+2N_B^{\free}\bigr)\times N_E^{\free}.
\end{equation}
A nontrivial solution of Eq.~\eqref{eq:generic_Cc0} then requires
\begin{equation}
\mathrm{rank}\,C(\epsilon)<N_E^{\free},
\end{equation}
which is equivalent to the simultaneous vanishing of all \(N_E^{\free}\times N_E^{\free}\) maximal minors of \(C(\epsilon)\).
Thus, unlike the square-matrix case, one is no longer tuning a single scalar condition with \(\epsilon\), but trying to satisfy several independent conditions at once.

This is why allowing \(\epsilon\) to vary does not restore generic solvability.
For a fixed sample, the spectral parameter \(\epsilon\) is indeed the quantity one normally tunes to search for bound states; however, in the open sector it is only one continuous parameter, while the bound-state condition requires the simultaneous disappearance of all forbidden amplitudes.
Accordingly, one does not generically expect any solution at all.
If a solution exists, it is an exceptional fine-tuned point rather than a stable feature.
Equivalently, in the full parameter space of boundary profiles, driving amplitudes, and sample parameters, the subset for which such a bound state exists is lower-dimensional and therefore nongeneric (measure zero).

A Floquet bound state in the continuum is precisely such an exceptional case.
Examples are known in specially engineered driven systems \cite{Della2014}, but in the open Floquet geometries considered here they require this kind of overdetermined matching; see also Refs.~\cite{Yajima1983,Kaneta1987}.
We therefore conclude that true bound trapping of a propagating branch is generically absent, and hence
\begin{equation}
p_\mu(\epsilon_0)=1
\end{equation}
for generic parameter values.

\begin{figure}[t]
    \centering
    \includegraphics[width=\linewidth]{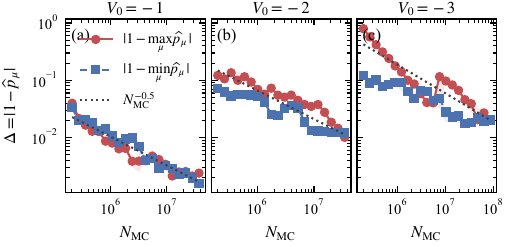}
    \caption{
    \textbf{Monte Carlo convergence of the branch weights toward generic openness.}
    Log--log plot of the extremal deviations
    \(\Delta_{\max}=|1-\max_\mu \widehat p_\mu^{(N_{\rm MC})}|\)
    and
    \(\Delta_{\min}=|1-\min_\mu \widehat p_\mu^{(N_{\rm MC})}|\)
    as functions of the Monte Carlo sample number \(N_{\rm MC}\).
    The data are computed for a sharply contacted Floquet lattice with
    \(V(x,t)=V\cos(2\pi x-\omega t)+V_0\),
    \(V=8\), \(\omega=1\), and \(V_0=-1,-2,-3\).
    The dotted reference line indicates the standard Monte Carlo scaling
    \(N_{\rm MC}^{-1/2}\).
    For all three offsets, both extremal deviations continue to decrease with this scaling and show no saturation plateau up to the largest sampled \(N_{\rm MC}\).
    A plateau in \(\log|1-p_\mu|\) would indicate convergence to a value different from \(p_\mu=1\); its absence supports the generic-openness conclusion that all propagating Floquet branches remain accessible and escapable.
    The slower convergence for more negative \(V_0\) is attributed to increasingly narrow resonant structures on the phase torus.
    }
    \label{fig:pmu_mc}
\end{figure}

To support this generic-openness conclusion numerically, we consider a sharply contacted Floquet lattice whose driven bulk region is described by
\[
V(x,t)=V\cos(2\pi x-\omega t)+V_0,
\qquad
V=8,\quad \omega=1,
\]
with three representative offsets \(V_0=-1,-2,-3\).
For each parameter set, we evaluate the Monte Carlo estimators
\(\widehat p_\mu^{(N_{\rm MC})}\)
of the branch weights and monitor the extremal deviations from unity,
\begin{equation}
\begin{split}
\Delta_{\max}(N_{\rm MC}) &= \left|1-\max_\mu \widehat p_\mu^{(N_{\rm MC})}\right|, \\
\Delta_{\min}(N_{\rm MC}) &= \left|1-\min_\mu \widehat p_\mu^{(N_{\rm MC})}\right|.
\end{split}
\label{eq:generic_mc_extrema}
\end{equation}
Figure~\ref{fig:pmu_mc} shows \(\log \Delta_{\max}\) and \(\log \Delta_{\min}\) versus \(\log N_{\rm MC}\).
For all three values of \(V_0\), both curves continue to decrease with the characteristic slope \(-1/2\), in agreement with the standard Monte Carlo scaling discussed in Appendix~\ref{appsubsec:mc_estimator}.
No plateau is observed up to \(N_{\rm MC}=10^8\).
This point is important: had \(p_\mu\) converged to a value different from \(1\), the corresponding \(\log |1-p_\mu|\) curve would eventually level off.
Instead, the absence of such a plateau and the persistent \(-1/2\) decay indicate that both the maximal and minimal branch weights continue to converge toward \(1\).

This behavior is especially noteworthy for \(V_0<0\).
In the corresponding static problem, lowering \(V_0\) would render branches with central frequency below zero inaccessible from scattering states and, correspondingly, unable to escape.
By contrast, in the Floquet problem these branches remain fully open to the scattering sector through sideband-assisted coupling, although the convergence of \(\widehat p_\mu^{(N_{\rm MC})}\) becomes visibly slower because of narrow resonant structures.
The numerical data therefore support the generic conclusion that propagating Floquet branches remain open, namely \(p_\mu=1\), even in parameter regimes where the analogous static system would exhibit inaccessible below-threshold branches.

\section{Boundary-robust long-sample transmission asymmetry}
\label{sec:robust_asymmetry}

We now turn to the robust long-sample observable implied by the open-sector theory.
The shrinking-window smoothing procedure was introduced in Sec.~\ref{sec:interface_scattering}.
Here we only use its physical consequence: for a finite driven segment, the pointwise transmission is dominated by dense Fabry--P\'erot oscillations, whereas the smoothed transmission removes this sample-specific interference while retaining the local spectral transport weight.
Thus the stable observable of the long-sample problem is not the detailed transmission line shape, but the smoothed transmission and the integrated quantities derived from it.

We use the quasienergy \(\epsilon\) as the fundamental variable.
The incident-channel index \(\alpha\) simultaneously encodes the laboratory energy and the incident lead, with
\[
E=\epsilon+n\omega ,
\qquad n\in\mathbb Z.
\]
Thus \((\alpha,\epsilon)\) labels a physical incident state.
The propagating modes inside the lattice are labeled by \((\mu,\epsilon)\), where \(\mu\) distinguishes Floquet--Bloch branches.

For a unit-current scattering state incident in channel \(\alpha\), the transmittance can be evaluated as the conserved current through any cross section.
We evaluate it in the deep bulk of the Floquet region.
There the scattering state decomposes as
\begin{equation}
\begin{split}
\Psi_{\alpha}^{(-),\cur}(x,t;\epsilon;L)
&=
\sum_{\mu\in\mathrm{prop}(\epsilon)}
d_{\mu\alpha}(\epsilon;L)\,
\Upsilon_\mu^{\cur}(x,t;\epsilon) \\
&\quad
+\eta_\alpha(x,t;\epsilon;L),
\end{split}
\label{eq:asym_deep_bulk_expansion}
\end{equation}
where the current-normalized branch \(\Upsilon_\mu^{\cur}\) carries current
\begin{equation}
j_\mu=\sgn(v_\mu),
\qquad
v_\mu=\frac{\partial\epsilon}{\partial k_\mu},
\label{eq:asym_branch_current}
\end{equation}
and the evanescent term \(\eta_\alpha\) is exponentially small at a deep-bulk observation point.
The coefficient \(d_{\mu\alpha}\) therefore directly measures how strongly the incident channel \(\alpha\) populates branch \(\mu\).

After shrinking-window smoothing and taking the long-sample limit, the branch intensity
\(|d_{\mu\alpha}|^2\) becomes the branch population \(p_{\mu\alpha}\).
The smoothed transmission for the incident channel \(\alpha\) is then the corresponding signed bulk current,
\begin{equation}
\bar{\mathcal T}_{\alpha}(\epsilon)
=
\sum_{\mu}
\sigma_\alpha\,p_{\mu\alpha}(\epsilon)\,\sgn(v_\mu),
\label{eq:asym_Talpha_branch_population}
\end{equation}
where
\[
\sigma_\alpha=+1
\quad
(\alpha\in\mathcal A_{\open}^{L}),
\qquad
\sigma_\alpha=-1
\quad
(\alpha\in\mathcal A_{\open}^{R}).
\]
The factor \(\sigma_\alpha\) converts the signed \(x\)-current into the positive transmittance in the corresponding incident direction.

Equation~\eqref{eq:asym_Talpha_branch_population} is the branch-population form of the smoothed transmittance.
It also makes clear why the response is local in incident energy.
The wave-function vectors entering boundary matching are frequency-local objects; hence, at fixed quasienergy, an incident channel efficiently populates only the nearby propagating branches to which it is connected by the boundary.
The channel-resolved weights \(p_{\mu\alpha}\) are therefore localized in the incident-channel index \(\alpha\).
The smoothed transmission is not a featureless average over all propagating branches, but a branch-selective current determined by the branches populated by the chosen incident channels.

We now choose an incident energy window \(\mathrm{Inc}\).
When the target Floquet band is sufficiently isolated in frequency, \(\mathrm{Inc}\) can be chosen so that the incident channels in this window predominantly populate the branches of a single target band \(B_{\rm tar}\).
Equivalently,
\[
\sum_{\alpha\in\mathrm{Inc}} p_{\mu\alpha}(\epsilon)
\approx 1
\quad
\text{for }
\mu\in B_{\rm tar},
\]
while the corresponding sum is negligible for branches outside \(B_{\rm tar}\).
Using the generic-openness result of Sec.~\ref{sec:generic_openness},
\[
p_\mu(\epsilon)
=
\sum_{\alpha\in\mathcal A_{\open}(\epsilon)}
p_{\mu\alpha}(\epsilon)
=
1
\]
for nonexceptional parameter values, the integrated left--right transmission asymmetry becomes
\begin{equation}
\begin{split}
\mathbf A_{\mathrm{Inc}}
&=
\int_{0}^{\hbar\omega}d\epsilon\,
\sum_{\alpha\in\mathrm{Inc},\mu}
\sgn(v_\mu)\,p_{\mu\alpha}(\epsilon)
\\
&\approx
\int_{0}^{\hbar\omega}d\epsilon\,
\sum_{\mu\in B_{\rm tar}}
\sgn(v_\mu).
\end{split}
\label{eq:asymmetry_integral}
\end{equation}
The final expression counts the net chirality, or equivalently the net crossing number, of the target band over one quasienergy Brillouin zone.
It is therefore fixed by the winding contribution \(C_{B_{\rm tar}}\) of that band:
\begin{equation}
\mathbf A_{\mathrm{Inc}}
\approx
C_{B_{\rm tar}}\,\hbar\omega .
\label{eq:asymmetry_plateau}
\end{equation}
With \(\hbar=1\), this is \(C_{B_{\rm tar}}\omega\).

This is the boundary-robust topological statement.
The detailed line shapes of
\(\bar{\mathcal T}^{L}(E)\) and
\(\bar{\mathcal T}^{R}(E)\) may be strongly reshaped by nonadiabatic boundaries, because the channel-resolved distribution \(p_{\mu\alpha}\) is boundary dependent.
However, the integrated asymmetry depends only on the total population exhausted on each propagating branch.
For generically open branches this total population is unity, so microscopic boundary redistribution drops out of the accumulated signal.
The plateau height is consequently locked to the bulk winding rather than to the detailed boundary profile.

The adiabatic boundary is therefore a clean benchmark for this mechanism, not its origin.
When the boundary varies slowly in space, branch mixing, backscattering, and sideband redistribution are suppressed, and the smoothed spectra resolve the connected Floquet--Bloch branches most directly.
In this limit the left- and right-incident transmission windows approach the rigid spectral shifts discussed in the companion article, so the asymmetry plateau can be read off transparently.
Away from the adiabatic limit, the spectra may be strongly distorted, but the integrated plateau persists as long as the same target band is selected and the propagating branches remain generically open.

\begin{figure}[ht]
    \centering
    \includegraphics[width=\columnwidth]{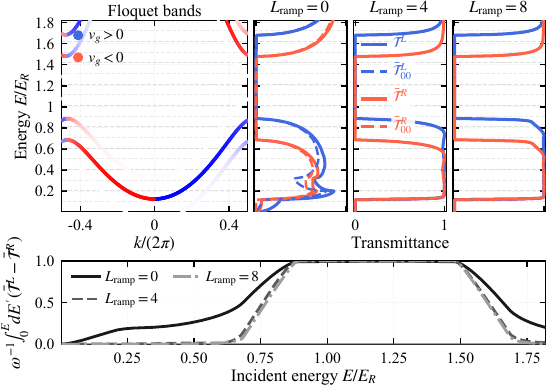}
    \caption{
    \textbf{Boundary robustness of the integrated spectral asymmetry.}
    For an open Floquet lattice with
    \(V(x,t)=\xi(x)V_{\rm bulk}(x,t)\),
    \(V_{\rm bulk}(x,t)=0.8E_R\cos(2\pi x-\omega t)+0.2E_R\),
    and
    \(E_R=\hbar^2\pi^2/(2md^2)\),
    the boundary envelope \(\xi(x)\) implements a linear ramp from \(0\) to \(1\) over a length \(L_{\rm ramp}\).
    Varying \(L_{\rm ramp}\) strongly distorts the smoothed transmission spectra
    \(\bar{\mathcal T}^{L/R}(E)\), while the integrated asymmetry
    \(\mathbf A(E_0)\) remains nearly invariant and saturates to the same plateau.
    The plateau height matches the rigid spectral shift in the adiabatic limit and is fixed by the winding of the selected bulk Floquet band.
    }
    
    \label{fig:spectral_asymmetry}
\end{figure}

Figure~\ref{fig:spectral_asymmetry} illustrates this point in a ramped-boundary geometry.
The driven bulk is held fixed, while only the boundary envelope is varied.
Changing the ramp length substantially reshapes the smoothed spectra
\(\bar{\mathcal T}^{L/R}(E)\), especially away from the adiabatic regime.
Nevertheless, the integrated asymmetry \(\mathbf A(E_0)\) remains nearly unchanged and saturates to the same plateau.
The plateau height agrees with the rigid-shift value in the adiabatic limit and is fixed by the winding of the selected Floquet band.

\section{Conclusion}
\label{sec:conclusion}

We have developed a scattering-state theory for open one-dimensional Floquet lattices based on a frequency-domain transfer-matrix formulation. The conjugate-symplectic structure of the transfer matrix organizes bulk Floquet--Bloch modes into propagating and evanescent sectors and provides the natural framework for interface matching, finite-length composition, and the shrinking-window smoothing required to define meaningful long-sample transport observables.

Within this framework, we formulated the open sector in a branch-resolved way by asking which propagating Floquet--Bloch branches are physically accessed deep inside a long sample by incoming scattering states. This led to branch-resolved weights \(p_{\mu\alpha}\) and total branch weights \(p_\mu\), and to the central identity that \(p_\mu\) equals the escape probability of a wave packet initialized on the corresponding bulk branch. This identification separates branch accessibility from branch openness and gives a direct physical interpretation of the long-sample branch weights.

We then showed that, in the open Floquet geometries considered here, true bound trapping of a propagating branch is nongeneric. Because a genuine bound state in an open quasienergy sector requires an overdetermined matching between decaying and propagating sectors, it is absent for generic parameter values. As a result, one generically has \(p_\mu=1\). This generic openness is the key structural input that connects the branch-resolved scattering theory to topological transport.

The main physical consequence is that long-sample transport is governed by deep-bulk branch populations rather than by boundary-sensitive interference details. After shrinking-window smoothing, the robust topological observable is not the detailed transmission line shape, which may be strongly reshaped by nonadiabatic boundaries, but the integrated left--right transmission asymmetry. When an incident-energy window selectively populates an isolated Floquet band, this asymmetry reduces to the net chirality, and hence the winding contribution, of that band. In this way, the boundary-robust asymmetry plateau emphasized in the companion article is placed on a full scattering-state foundation.

Finally, we clarified the role of spatially adiabatic boundaries. They provide the clearest spectroscopic benchmark for exposing the underlying branch structure by suppressing unnecessary branch mixing and backscattering, but they are not the origin of the topological response. The topological content lies instead in the robustness of the integrated asymmetry itself, which survives substantial boundary deformation.

More broadly, the framework developed here provides a systematic route from microscopic scattering in open Floquet devices to robust bulk-sensitive transport observables. We expect it to be useful for analyzing other open driven systems in which boundary-induced mode conversion obscures the underlying Floquet topology at the level of detailed transmission spectra.

\section*{Acknowledgement}

\appendix

\section{Algebraic structure of conjugate-symplectic transfer matrices}
\label{app:tm_algebraic}

\subsection{Exponential generation in the diagonalizable case}
\label{appsubsec:tm_expgen}

We justify here the reduction from finite-step transfer matrices \(F\) to infinitesimal generators \(G\).

\begin{theorem}[Detailed exponential generation]
\label{thm:tm_expgen_detailed}
Let \(F\) satisfy \(F^\dagger JF=J\) and suppose \(F\) is diagonalizable:
\[
F=S D_F S^{-1}.
\]
Choose a branch of the complex logarithm for each eigenvalue and define
\[
D_G=\diag(g_1,\dots,g_{2N}),
\qquad
e^{D_G}=D_F,
\]
with the branch choice made so that partner indices satisfy
\begin{equation}
g_{p(j)}=-g_j^*,
\label{eq:app_tm_log_pairing}
\end{equation}
where \(\lambda_{p(j)}=1/\lambda_j^*\).
Define
\[
G\equiv S D_G S^{-1}.
\]
Then
\[
e^G=F,
\qquad
G^\dagger J+JG=0.
\]
\end{theorem}

\begin{proof}
The identity \(e^G=F\) is immediate:
\[
e^G=e^{S D_G S^{-1}}=S e^{D_G} S^{-1}=S D_F S^{-1}=F.
\]

To verify the infinitesimal conjugate-symplectic relation, define
\[
K\equiv S^\dagger J S.
\]
From \(F^\dagger JF=J\) and \(F=S D_F S^{-1}\), one finds
\[
D_F^\dagger K D_F=K.
\]
Since \(D_F\) is diagonal, the \((i,j)\)-entry gives
\[
\lambda_i^*\lambda_j\,K_{ij}=K_{ij}.
\]
Hence \(K_{ij}\neq0\) only if \(\lambda_i^*\lambda_j=1\), namely only within reciprocal-conjugate spectral classes.

Now compute
\[
G^\dagger J+JG
=
(S^{-1})^\dagger\bigl(D_G^\dagger K + K D_G\bigr)S^{-1}.
\]
Its \((i,j)\)-entry is
\[
(D_G^\dagger K + K D_G)_{ij}=(g_i^*+g_j)K_{ij}.
\]
If \(K_{ij}=0\), the entry vanishes trivially.
If \(K_{ij}\neq0\), then \(j=p(i)\), and by Eq.~\eqref{eq:app_tm_log_pairing},
\[
g_j=-g_i^*.
\]
Thus every entry vanishes and therefore
\[
D_G^\dagger K + K D_G=0.
\]
It follows that
\[
G^\dagger J+JG=0.
\]
\end{proof}

\begin{remark}
Under \(G^\dagger J+JG=0\), the generator necessarily has the block form
\[
G=
\begin{bmatrix}
A&B\\
C&-A^\dagger
\end{bmatrix},
\qquad
B^\dagger=B,\quad C^\dagger=C.
\]
This is the form used in the proof of the sign theorem below.
\end{remark}

\subsection{Reciprocal-conjugate pairing and \(J\)-orthogonality}
\label{appsubsec:tm_pairing}

\begin{proof}[Proof of Lemma~\ref{lem:tm_pairing}]
From \(F^\dagger JF=J\), one obtains
\[
F^{-1}=J^{-1}F^\dagger J=-J F^\dagger J,
\]
hence
\[
F=-J F^{-\dagger}J.
\]
Therefore \(F\) and \(F^{-\dagger}\) are similar, and so they have the same spectrum.
But the spectrum of \(F^{-\dagger}\) is
\[
\sigma(F^{-\dagger})=\left\{\frac{1}{\lambda^*}:\lambda\in\sigma(F)\right\}.
\]
Hence
\[
\lambda\in\sigma(F)\quad\Longrightarrow\quad \frac{1}{\lambda^*}\in\sigma(F),
\]
which proves reciprocal-conjugate pairing.
\end{proof}

\begin{proof}[Proof of Lemma~\ref{lem:tm_Jorth}]
Using \(F^\dagger JF=J\), one has
\[
y_m^\dagger J y_n
=
y_m^\dagger F^\dagger JF y_n
=
\lambda_m^*\lambda_n\, y_m^\dagger J y_n.
\]
Hence
\[
(1-\lambda_m^*\lambda_n)\,y_m^\dagger J y_n=0.
\]
Whenever \(\lambda_m^*\lambda_n\neq1\), this implies
\[
y_m^\dagger J y_n=0.
\]
\end{proof}

\subsection{Perturbation theory}\label{appsubsec:perturbation}
Let \(Q(u)\) be a smooth one-parameter family of finite-dimensional CSp operators (or generators) depending on a real parameter \(u\), and suppose \(Q(0)=Q\) is diagonalizable. We consider a small perturbation
\[
Q(u) = Q + uW + o(u),
\]
with \(W\) a fixed matrix (not necessarily CSp). Denote an eigenpair of \(Q\) by
\[
Q Y = \lambda Y,
\]
where \(Y\neq 0\). Let \(Y^\ddagger\) denote a left eigenvector (row vector) satisfying \(Y^\ddagger Q = \lambda Y^\ddagger\) and normalized so that \(Y^\ddagger Y = 1\) (such a biorthonormal pair exists for diagonalizable \(Q\) after appropriate local normalization; see main supplement).

\paragraph{(i) Simple eigenvalue (non-degenerate).}
Consider the perturbed eigenpair \((\lambda(u),Y(u))\) with \(\lambda(0)=\lambda\), \(Y(0)=Y\). Expand
\[
(Q + uW)\bigl(Y + u\dot Y\bigr) + o(u) = \bigl(\lambda + u\dot\lambda\bigr) \bigl(Y + u\dot Y\bigr) + o(u),
\]
where \(\dot\lambda=\left.\dfrac{d\lambda}{du}\right|_{u=0}\), \(\dot Y=\left.\dfrac{dY}{du}\right|_{u=0}\). Collect \(O(u)\) terms:
\begin{equation}\label{eq:Ou}
W Y + (Q-\lambda)\dot Y = \dot\lambda\, Y.
\end{equation}
Left-multiply by the dual row \(Y^\ddagger\). Using \(Y^\ddagger (Q-\lambda)=0\) we obtain
\begin{equation}\label{eq:pert_simple}
\dot\lambda \;=\; Y^\ddagger W Y .
\end{equation}
This is the standard first-order perturbation formula for a simple eigenvalue.

\paragraph{(ii) Geometrically degenerate eigenvalue (multiplicity \(m>1\)).}
Assume \(\lambda\) has geometric multiplicity \(m\). Choose a biorthonormal basis \(\{Y_1,\dots,Y_m\}\) of the \(\lambda\)-eigenspace and corresponding left vectors \(\{Y_1^\ddagger,\dots,Y_m^\ddagger\}\) such that \(Y_i^\ddagger Y_j=\delta_{ij}\) (possible under diagonalizability; see main supplement for constructive procedure). Expand a perturbed eigenvector inside this subspace:
\[
Y(u) = \sum_{j=1}^m c_j(u)\, Y_j + Z(u),
\qquad Z(u)\perp\!\!\perp \text{(eigenspace)}.
\]
Insert into \((Q+uW)Y(u) = \lambda(u) Y(u)\) and collect \(O(u)\) terms, then project onto the \(i\)-th dual row \(Y_i^\ddagger\). Using \(Y_i^\ddagger (Q-\lambda)=0\) and \(Y_i^\ddagger Z(0)=0\), we obtain the \(m\times m\) secular (matrix) equation:
\begin{equation}\label{eq:degenerate_matrix}
\sum_{j=1}^m \bigl( Y_i^\ddagger W Y_j \bigr) c_j(0) \;=\; \dot\lambda\, c_i(0),\qquad i=1,\dots,m.
\end{equation}
In matrix form \(M\mathbf c = \dot\lambda\,\mathbf c\) with \(M_{ij}=Y_i^\ddagger W Y_j\). Thus the admissible first-order splitting rates \(\dot\lambda\) are the eigenvalues of the finite matrix \(M\); diagonalizing \(M\) yields the linear splitting directions and rates. In particular degeneracy is generically broken linearly.

\subsection{Proof of the sign equal division theorem}
\label{appsubsec:tm_sign}

We now prove Theorem~\ref{thm:tm_sign}.
By Theorem~\ref{thm:tm_expgen_detailed} and \ref{appsubsec:perturbation}, it suffices to analyze a generator of the form
\[
G=
\begin{bmatrix}
A&B\\
C&-A^\dagger
\end{bmatrix},
\qquad
B^\dagger=B,\quad C^\dagger=C.
\]
\begin{proof}[Proof of Theorem~\ref{thm:tm_sign}]
Consider the polynomial
\begin{equation}
\begin{aligned}
p(i\lambda,u)
&=\det(G-\lambda-uJ)
\\
&=\det\begin{bmatrix}
    A-\lambda I & B-uI\\
    C+uI & -A^\dagger-\lambda I
\end{bmatrix},
\end{aligned}
\label{eq:app_sign_poly_1}
\end{equation}
where \(u\) is real and \(\lambda\) is imaginary. 
We are interested in the imaginary-\(\lambda\) solutions of
\(p(i\lambda,0)=0\). 
These solutions correspond to propagating eigenvectors with
\(iY^\dagger JY\neq0\).

We now perturb the roots by varying \(u\).
For a simple root, first-order perturbation theory gives
\begin{equation}
(Y^\dagger JY)\,
\left.\frac{d(i\lambda)}{du}\right|_{u=0}
=
iY^\dagger JWY .
\label{eq:app_sign_perturb}
\end{equation}
Setting \(W=-J\), we obtain
\begin{equation}
\left.\frac{du}{d(i\lambda)}\right|_{u=0}
=
\frac{Y^\dagger JY}{iY^\dagger Y}
\in[-1,1].
\label{eq:app_sign_slope}
\end{equation}
Thus the current sign is encoded in the slope of the root curve
\(p(i\lambda,u)=0\) at \(u=0\):
\begin{equation}
\sgn(-iY^\dagger JY)
=
\sgn\!\left[
\left.\frac{du}{d(i\lambda)}\right|_{u=0}
\right].
\label{eq:app_sign_current_slope}
\end{equation}

\begin{figure}[ht]
    \centering
    \includegraphics[width=1\linewidth]{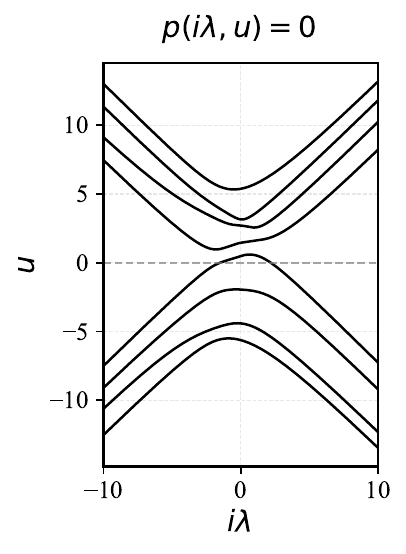}
    \caption{Diagram of the root curves \(p(i\lambda,u)=0\).}
    \label{fig:ilambdau}
\end{figure}

Next consider the asymptotic regime \(|i\lambda|\to\infty\).
For the polynomial to vanish, one must have
\[
\frac{|u|}{|\lambda|}\to1 .
\]
Therefore each root curve has one of four possible asymptotic types, classified by the signs of \(u\) as
\(i\lambda\to-\infty\) and \(i\lambda\to+\infty\).
We label these four types as
\[
1:\ (-,+),\quad
2:\ (+,+),\quad
3:\ (+,-),\quad
4:\ (-,-),
\]
where the first sign refers to \(i\lambda\to-\infty\) and the second sign to \(i\lambda\to+\infty\).
These four possibilities are shown in Fig.~\ref{fig:four_types}.

\begin{figure}[ht]
    \centering
    \includegraphics[width=1\linewidth]{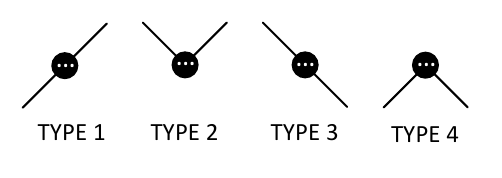}
    \caption{Four possible asymptotic types of root curves in the \((i\lambda,u)\) plane. 
    The labels \(1,2,3,4\) correspond respectively to the sign pairs
    \((-+)\), \((++)\), \((+-)\), and \((--)\), where the two signs denote the sign of \(u\) as \(i\lambda\to-\infty\) and \(i\lambda\to+\infty\).}
    \label{fig:four_types}
\end{figure}

Let \(N_+\) and \(N_-\) denote the numbers of roots at \(u=0\) with
\[
\frac{du}{d(i\lambda)}>0,
\qquad
\frac{du}{d(i\lambda)}<0,
\]
respectively.
Crossings of the \(u=0\) axis with positive slope occur for type-1 curves, while crossings with negative slope occur for type-3 curves.
Type-2 and type-4 curves contribute an equal number of positive- and negative-slope crossings, and hence do not contribute to the difference.
Therefore
\begin{equation}
N_+-N_-=N_1-N_3,
\label{eq:app_sign_Ndiff}
\end{equation}
where \(N_j\) is the number of curves of type \(j\).

It remains to show that \(N_1=N_3\).
For this purpose, rewrite the same polynomial as
\begin{equation}
\begin{aligned}
p(i\lambda,u)
&=
\det\begin{bmatrix}
    B-uI & -A+\lambda I\\
    -A^\dagger-\lambda I & -C-uI
\end{bmatrix}
\\
&=
\det[H(\lambda)-uI].
\end{aligned}
\label{eq:app_sign_poly_2}
\end{equation}
For imaginary \(\lambda\), \(H(\lambda)\) is Hermitian, so the roots
\(u_i(\lambda)\) are real.
Moreover,
\begin{equation}
\mathrm{Tr}\,H(\lambda)=\sum_i u_i(\lambda)
\label{eq:app_sign_trace}
\end{equation}
is independent of \(\lambda\).
Consequently,
\begin{equation}
\lim_{i\lambda\rightarrow +\infty}
\frac{\sum_i u_i(\lambda)}{|i\lambda|}
=
\lim_{i\lambda\rightarrow -\infty}
\frac{\sum_i u_i(\lambda)}{|i\lambda|}
=0.
\label{eq:app_sign_trace_limit}
\end{equation}
On the other hand, the four asymptotic curve types imply
\begin{equation}
\lim_{i\lambda\rightarrow +\infty}
\frac{\sum_i u_i(\lambda)}{|i\lambda|}
-
\lim_{i\lambda\rightarrow -\infty}
\frac{\sum_i u_i(\lambda)}{|i\lambda|}
=
2(N_1-N_3).
\label{eq:app_sign_N1N3}
\end{equation}
Combining Eqs.~\eqref{eq:app_sign_trace_limit} and
\eqref{eq:app_sign_N1N3} gives
\[
N_1=N_3.
\]
Therefore \(N_+=N_-\), proving the current-sign equal division theorem.
\end{proof}

\section{Dual basis and normalization formulas}
\label{app:tm_dualnorm}

\subsection{Dual-basis construction in the diagonalizable case}
\label{appsubsec:tm_dual}

We now supply the construction behind Proposition~\ref{prop:tm_biorth}.

\begin{proof}[Proof of Proposition~\ref{prop:tm_biorth}]
Let
\[
F=S D_F S^{-1},
\qquad
S=[Y_1,\dots,Y_{2N}],
\qquad
N\equiv S^\dagger J S.
\]
By Lemma~\ref{lem:tm_Jorth}, \(N_{mn}=0\) whenever \(\lambda_m^*\lambda_n\neq1\).
Hence \(N\) is block-sparse with respect to reciprocal-conjugate spectral classes.

For each reciprocal-conjugate class \(\mathcal C\), let \(S_{\mathcal C}\) denote the matrix formed by the corresponding eigenvectors, and let
\[
N_{\mathcal C}=S_{\mathcal C}^\dagger J S_{\mathcal C}
\]
be the local \(J\)-Gram matrix.
After possible linear recombination inside degenerate self-paired classes on the unit circle, we may assume that each \(N_{\mathcal C}\) is invertible.
Then define the local dual rows by
\begin{equation}
S_{\mathcal C}^{\ddagger}
=
N_{\mathcal C}^{-1}S_{\mathcal C}^\dagger J.
\label{eq:app_tm_local_dual}
\end{equation}
By construction,
\[
S_{\mathcal C}^{\ddagger}S_{\mathcal C}
=
N_{\mathcal C}^{-1}S_{\mathcal C}^\dagger J S_{\mathcal C}
=
I.
\]

Two cases are worth recording explicitly.

\paragraph{Paired class.}
Suppose \(\mathcal C\) consists of two distinct reciprocal-conjugate eigenvalues \(\lambda\) and \(1/\lambda^*\), each of multiplicity \(r\).
Order the corresponding eigenvectors as
\[
S_{\mathcal C}=[\,U\mid V\,],
\qquad
U=[u_1,\dots,u_r],\quad V=[v_1,\dots,v_r],
\]
and define
\[
M=V^\dagger J U.
\]
Then
\begin{equation}
N_{\mathcal C}
=
S_{\mathcal C}^\dagger J S_{\mathcal C}
=
\begin{bmatrix}
0&-M^\dagger\\
M&0
\end{bmatrix}.
\label{eq:app_tm_paired_gram}
\end{equation}
If \(M\) is invertible, the local dual is given directly by Eq.~\eqref{eq:app_tm_local_dual}.

\paragraph{Self-paired class on the unit circle.}
If \(|\lambda|=1\), let \(W\) collect the eigenvectors of that class.
Then
\[
N_{\mathcal C}=W^\dagger J W
\]
is skew-Hermitian.
After a unitary recombination inside the class one may diagonalize it as
\[
N_{\mathcal C}=U D U^\dagger,
\]
with \(D\) purely imaginary diagonal and nonsingular in the regular case.
Setting
\[
W_{\mathrm{diag}}=WU
\]
gives
\[
W_{\mathrm{diag}}^\dagger J W_{\mathrm{diag}}=D,
\]
and Eq.~\eqref{eq:app_tm_local_dual} again yields the local dual rows.

\paragraph{Assembly.}
Assembling all local dual blocks in the same global ordering gives a matrix \(S^{\ddagger}\) such that
\[
S^{\ddagger}S=I_{2N}.
\]
Since \(S\) is invertible, it also follows that
\[
SS^{\ddagger}=I_{2N}.
\]
Thus the individual columns \(\tilde y_\alpha\) of \(S\) and rows \(\tilde y_\alpha^{\ddagger}\) of \(S^{\ddagger}\) satisfy
\[
\tilde y_\alpha^{\ddagger}\tilde y_\beta=\delta_{\alpha\beta},
\qquad
\sum_{\alpha=1}^{2N}\tilde y_\alpha\tilde y_\alpha^{\ddagger}=I_{2N},
\]
which is the desired \(J\)-biorthogonal completeness relation.
\end{proof}

\subsection{Proof of \(\rho_\mu v_\mu=j_\mu\)}
\label{appsubsec:tm_nvj}

\begin{proof}[Proof of Proposition~\ref{prop:tm_nvj}]
Let
\[
A(x,\epsilon)=F(x,0;\epsilon),
\qquad
B(x,\epsilon)=F(1,x;\epsilon),
\]
so that
\[
F(\epsilon)=B(x,\epsilon)A(x,\epsilon).
\]
Since \(\partial_\epsilon M(x,\epsilon)=-I\),
\[
\partial_\epsilon G(x,\epsilon)=
\begin{bmatrix}
0&0\\
-I&0
\end{bmatrix}.
\]
Differentiating the path-ordered exponential gives
\begin{equation}
\partial_\epsilon F(\epsilon)
=
\int_0^1
B(x,\epsilon)
\begin{bmatrix}
0&0\\
-I&0
\end{bmatrix}
A(x,\epsilon)\,dx.
\label{eq:app_tm_dFdE}
\end{equation}

Let \(Fy_\mu=\lambda_\mu y_\mu\), with \(\lambda_\mu=e^{ik_\mu}\) on the unit circle.
Differentiate the eigenvalue equation:
\[
(\partial_\epsilon F)y_\mu + F\,\partial_\epsilon y_\mu
=
(\partial_\epsilon\lambda_\mu)y_\mu + \lambda_\mu\,\partial_\epsilon y_\mu.
\]
Left-multiplying by \(y_\mu^\dagger JF^{-1}\), and using
\[
y_\mu^\dagger JF^{-1}=\lambda_\mu^*\,y_\mu^\dagger J,
\]
one finds
\begin{equation}
\frac{\partial_\epsilon\lambda_\mu}{\lambda_\mu}
=
\frac{y_\mu^\dagger JF^{-1}(\partial_\epsilon F)y_\mu}{y_\mu^\dagger Jy_\mu}.
\label{eq:app_tm_dloglambda}
\end{equation}

Now use Eq.~\eqref{eq:app_tm_dFdE}.
Since \(F^{-1}B(x,\epsilon)=A(x,\epsilon)^{-1}\), one gets
\[
\begin{aligned}
y_\mu^\dagger JF^{-1}(\partial_\epsilon F)y_\mu
&= \int_0^1 y_\mu^\dagger J A^{-1}(x,\epsilon)
   \begin{bmatrix} 0 & 0 \\ -I & 0 \end{bmatrix} \\
&\quad A(x,\epsilon)y_\mu \, dx.
\end{aligned}
\]
Using the conjugate-symplectic identity \(A^\dagger J A=J\), equivalently \(J A^{-1}=A^\dagger J\), this becomes
\[
\int_0^1
Y_\mu(x,\epsilon)^\dagger
J
\begin{bmatrix}
0&0\\
-I&0
\end{bmatrix}
Y_\mu(x,\epsilon)\,dx.
\]
Writing \(Y_\mu=[\Phi_\mu,\Phi_\mu']^T\), the integrand is
\[
Y_\mu^\dagger
J
\begin{bmatrix}
0&0\\
-I&0
\end{bmatrix}
Y_\mu
=
-\Phi_\mu^\dagger\Phi_\mu.
\]
Therefore
\begin{equation}
y_\mu^\dagger JF^{-1}(\partial_\epsilon F)y_\mu
=
-\,\rho_\mu(\epsilon).
\label{eq:app_tm_num}
\end{equation}

Combining Eqs.~\eqref{eq:app_tm_dloglambda} and \eqref{eq:app_tm_num},
\[
\frac{\partial_\epsilon\lambda_\mu}{\lambda_\mu}
=
-\frac{\rho_\mu(\epsilon)}{y_\mu^\dagger Jy_\mu}.
\]
Since \(\lambda_\mu=e^{ik_\mu}\),
\[
\frac{\partial_\epsilon\lambda_\mu}{\lambda_\mu}
=
i\,\partial_\epsilon k_\mu
=
\frac{i}{v_\mu(\epsilon)}.
\]
Using \(j_\mu=-i\,y_\mu^\dagger Jy_\mu\), we obtain
\[
\rho_\mu(\epsilon)\,v_\mu(\epsilon)=j_\mu(\epsilon).
\]
\end{proof}
\subsection{Proof of current normalization versus spectral normalization}
\label{appsubsec:tm_currentnorm}

\begin{proof}[Proof of Proposition~\ref{prop:tm_current_en}]
Since
\[
\delta(k-k')=|v_\mu(\epsilon)|\,\delta(\epsilon-\epsilon'),
\]
Eq.~\eqref{eq:tm_energy_norm} gives quasienergy normalization from the \(k\)-normalized state \(\Phi_{\mu k}\).

For \(\Phi_{\mu k}(x,t)=(2\pi)^{-1/2}e^{ikx}u_{\mu k}(x,t)\), the one-cell probability is
\[
\rho_\mu(k)=\frac{1}{2\pi}.
\]
By Proposition~\ref{prop:tm_nvj},
\[
j_\mu(k)=\rho_\mu(k)v_\mu(k)=\frac{v_\mu(k)}{2\pi}.
\]
Multiplying the state by \(\sqrt{2\pi/|v_\mu|}\) therefore produces current
\[
\sgn(v_\mu),
\]
hence unit current magnitude.
Equivalently,
\[
\Upsilon_\mu^{\cur}(\epsilon)
=
\sqrt{\frac{2\pi}{|v_\mu(\epsilon)|}}\,
\Phi_{\mu,k_\mu(\epsilon)}
\]
is current normalized.
After a phase choice, this is exactly the unit-current normalization used in the transfer-matrix language.
\end{proof}

\section{Shrinking-window smoothing and nonstationary averaging}
\label{app:smoothing}

\subsection{Admissible energy envelopes and path-expansion representations}
\label{appsubsec:smoothing_setup}

We collect here the assumptions underlying the shrinking-window smoothing used in the main text.

For each sample length \(L\), let \(W(\epsilon;\epsilon_0)\) be a normalized energy-envelope amplitude centered at \(\epsilon_0\), satisfying
\[
\int d\epsilon\, |W(\epsilon;\epsilon_0)|^2 = 1,
\]
and localized in a quasienergy window of width \(\Delta\epsilon(L)\) with
\[
\Delta\epsilon(L)\to0,
\qquad
L\,\Delta\epsilon(L)\to\infty.
\]
Only \(|W|^2\) enters the averaged observables, so the phase of \(W\) is irrelevant for the smoothing procedure.

Whenever a path expansion is used, we assume an admissible representation in which all \(O(L)\) quasienergy oscillations are carried explicitly by the propagation matrices \(D_\oplus^L\) and \(D_\ominus^{-L}\).
Equivalently, after factoring out those explicit propagation phases, the product of \(|W(\epsilon;\epsilon_0)|^2\) with all remaining prefactors varies only on \(O(1)\) quasienergy scales across the shrinking window.
The final smoothed observables are representation independent; the admissible representation is used only to make the fast/slow separation explicit.

\subsection{Path expansion of the finite-length transmission amplitude}
\label{appsubsec:smoothing_path}

We now derive the delay-sector form of the smoothed transmittance.

Starting from Eq.~\eqref{eq:scatt_three_segment}, each matrix element \(T_{\beta\alpha}(\epsilon;L)\) is a sum over scattering paths \(s\) connecting the incoming lead channel \(\alpha\) to the outgoing lead channel \(\beta\).
Each path is specified by the sequence of propagating bulk branches visited between successive interface reflections.
We may therefore write
\begin{equation}
T_{\beta\alpha}(\epsilon;L)
=
\sum_{s\in\mathcal S_{\beta\alpha}}
A_{\beta\alpha}^{s}(\epsilon)\,
e^{i\phi_{\beta\alpha}^{s}(\epsilon;L)},
\label{eq:app_smoothing_path_expansion}
\end{equation}
where \(A_{\beta\alpha}^{s}(\epsilon)\) is the slowly varying path amplitude and \(\phi_{\beta\alpha}^{s}(\epsilon;L)\) is the total propagation phase accumulated along the long propagating segments of that path.

For a path with counting vector
\[
\mathbf N=
(N_{\mu_1}^+,\dots,N_{\mu_{N_B}}^+,N_{\nu_1}^-,\dots,N_{\nu_{N_B}}^-),
\]
the phase has the form
\begin{equation}
\begin{split}
& \phi_{\beta\alpha}^{s}(\epsilon;L) \\
& \quad = L \biggl( 
   \sum_j N_{\mu_j}^+ k_{\mu_j}^+(\epsilon)
   - \sum_j N_{\nu_j}^- k_{\nu_j}^-(\epsilon)
   \biggr). \\
\end{split}
\label{eq:app_smoothing_phase}
\end{equation}
Differentiating with respect to \(\epsilon\), we obtain the total Wigner delay
\begin{equation}
\begin{split}
& \tau_w^{s}(\epsilon;L) 
= \partial_\epsilon \phi_{\beta\alpha}^{s}(\epsilon;L) \\
& \quad = L \biggl( 
   \sum_j N_{\mu_j}^+ \frac{1}{v_{\mu_j}^+(\epsilon)}
   - \sum_j N_{\nu_j}^- \frac{1}{v_{\nu_j}^-(\epsilon)}
   \biggr).
\end{split}
\label{eq:app_smoothing_delay}
\end{equation}
For fixed path data, \(\tau_w^{s}\) is generically of order \(L\).

Substituting Eq.~\eqref{eq:app_smoothing_path_expansion} into Eq.~\eqref{eq:scatt_avg_Tba}, we obtain
\begin{equation}
\begin{split}
\avg{\mathcal T_{\beta\alpha}}^{(L)}(\epsilon_0)
&= \int d\epsilon\,
   |W(\epsilon;\epsilon_0)|^2\,
   |T_{\beta\alpha}(\epsilon;L)|^2
   \frac{|j_\beta(\epsilon)|}{|j_\alpha(\epsilon)|} \\
&= \int d\epsilon\,
   |W(\epsilon;\epsilon_0)|^2\,
   \frac{|j_\beta(\epsilon)|}{|j_\alpha(\epsilon)|} \\
&\quad \times \sum_{s,s'}
   A_{\beta\alpha}^{s}(\epsilon)
   A_{\beta\alpha}^{s'}(\epsilon)^*\,
   e^{i\Phi_{ss'}(\epsilon;L)}.
\end{split}
\label{eq:app_smoothing_double_sum}
\end{equation}
where
\[
\Phi_{ss'}(\epsilon;L)
=
\phi_{\beta\alpha}^{s}(\epsilon;L)-\phi_{\beta\alpha}^{s'}(\epsilon;L).
\]

If
\[
\partial_\epsilon\Phi_{ss'}(\epsilon;L)
=
\tau_w^{s}(\epsilon;L)-\tau_w^{s'}(\epsilon;L)
\]
is nonzero and of order \(L\), then the corresponding averaged cross term vanishes, because the bound
\[
\partial_\epsilon\bigl(|W|^2\bigr) < C\,\Delta\epsilon^{-2}
\]
ensures that the envelope \(|W|^2\) varies slowly and therefore contains no compensating \(O(L)\) Fourier components.
Hence only path pairs with equal total Wigner delay survive the shrinking-window average.
Regrouping the surviving contributions yields Eq.~\eqref{eq:scatt_delay_sector}.

If, in addition, equal delay implies equal counting vector \(\mathbf N\), then \(\phi_{\beta\alpha}^{s}(\epsilon_0;L)\) is common inside a given \(\mathcal S_{\mathbf N}\), so it factors out of the inner sum and disappears after taking the absolute square.
This yields Eq.~\eqref{eq:scatt_counting_vector}.

Finally, since the surviving contributions depend only on the local values of the slow factors at \(\epsilon_0\), the long-sample limit is independent of the detailed admissible choice of \(W\).

\subsection{The same path logic for branch-resolved bulk amplitudes}
\label{appsubsec:smoothing_bulk_path}

The later branch-resolved bulk quantities are obtained by terminating the scattering path not on an outgoing lead channel \(\beta\), but on a propagating bulk branch \(\mu\) inside the central homogeneous region.
Accordingly, one writes the deep-bulk branch amplitude as
\begin{equation}
d_{\mu\alpha}(\epsilon;L)
=
\sum_{s\in\mathcal S_{\mu\alpha}^{\mathrm{bulk}}}
B_{\mu\alpha}^{s}(\epsilon)\,
e^{i\varphi_{\mu\alpha}^{s}(\epsilon;L)},
\label{eq:app_smoothing_bulk_path}
\end{equation}
with exactly the same separation between slowly varying amplitudes and \(O(L)\) propagation phases.

For example, for left incidence from channel \(\alpha\) into a right-going bulk branch \(\mu\), the direct contribution is the corresponding entry of the left interface transmission block,
\[
d_{\mu\alpha}^{(0)}\sim (T_1)_{\mu\alpha},
\]
while the first round-trip correction has the schematic form
\[
d_{\mu\alpha}^{(1)}
\sim
\bigl(
R_3 D_\ominus^{-L} R_2 D_\oplus^{L} T_1
\bigr)_{\mu\alpha},
\]
up to local \(O(1)\) phases that may be absorbed into the centered bulk-branch convention used to define the deep-bulk basis.
Thus the path logic for the branch-resolved intensity
\[
\mathcal I_{\mu\alpha}(\epsilon;L)=|d_{\mu\alpha}(\epsilon;L)|^2
\]
is completely parallel to that for the transmission probability \(\mathcal T_{\beta\alpha}(\epsilon;L)\):
one merely replaces the terminal lead channel by a terminal bulk branch.

\section{Adiabatic boundary: full-channel local matching and long-boundary dephasing}
\label{app:adiabatic_boundary}

In the main text, the adiabatic boundary was described in the physical language of local Floquet--Bloch branch continuation through a slowly varying periodic structure.
The purpose of the present appendix is to formulate that picture in a technically controllable way.

The key point is that the local matching problem between two nearby structures must first be posed in the \emph{full} channel space.
In particular, near a band edge, propagating and evanescent partner modes must be treated together.
The connected-branch transmission used in the main text is then extracted from this full-channel matching problem.
The long-boundary conclusion follows by combining the local connected-transmission estimate with the same shrinking-window dephasing mechanism used elsewhere in the paper.

To regularize the smooth boundary, we discretize the slow parameter profile into a sequence of long homogeneous segments.
Each segment carries its own local Floquet problem, and neighboring segments differ only by \(O(\delta\xi)\).
The continuous connected-branch picture of the main text is recovered in the limit of fine discretization.

\subsection{Discrete regularization of a smooth boundary}
\label{appsubsec:adiabatic_discretization}

Let \(\xi\in[0,1]\) parametrize a smooth interpolation between the asymptotic lead side and the final bulk Floquet lattice.
We choose a partition
\begin{equation}
0=\xi_0<\xi_1<\cdots<\xi_{N_s}=1,
\qquad
\delta\xi_j=\xi_{j+1}-\xi_j,
\label{eq:app_adiabatic_partition}
\end{equation}
with
\[
\max_j \delta\xi_j = O(N_s^{-1}).
\]
Each interval is replaced by a long homogeneous segment of length \(L_0\), carrying the local Floquet lattice at parameter value \(\xi_j\).
The adiabatic limit is taken as
\begin{equation}
L_0\to\infty
\quad\text{followed by}\quad
N_s\to\infty.
\label{eq:app_adiabatic_limit_order}
\end{equation}

For fixed quasienergy \(\epsilon\), let
\[
X(\xi,\epsilon)
=
[\,X_\oplus(\xi,\epsilon)\mid X_\ominus(\xi,\epsilon)\,]
\]
denote a full local channel basis, including both propagating and evanescent modes.
Here \(X_\oplus\) and \(X_\ominus\) are the full \(+\) and \(-\) sectors in the sense used in the main text, not merely the propagating subspaces.
The local scattering problem between neighboring segments \(\xi_j\) and \(\xi_{j+1}\) is formulated in this full basis.
This is essential near band edges, where propagating and evanescent partner modes cannot be separated before the matching problem is solved.

\subsection{Gauge and scaling choices for regular channels}
\label{appsubsec:adiabatic_gauge}

We now explain the local gauge choices used in the matching analysis.
On a regular interval of the parameter path, the relevant propagating branches remain simple and separated, and the regular evanescent partner families may likewise be chosen continuously.

For the right-going propagating branches, we write
\[
B_\oplus(\xi,\epsilon)
=
[\,Y_1^+(\xi,\epsilon),\dots,Y_{N_B}^+(\xi,\epsilon)\,],
\]
with dual rows \(B_\oplus(\xi,\epsilon)^\ddagger\).
Because each propagating branch is simple on the regular interval, one may choose its phase continuously so as to satisfy the \(\xi\)-parallel transport condition
\begin{equation}
Y_\mu^+(\xi,\epsilon)^\ddagger\,\partial_\xi Y_\mu^+(\xi,\epsilon)=0.
\label{eq:app_adiabatic_parallel}
\end{equation}
Indeed, under the phase change
\[
Y_\mu^+\mapsto e^{i\vartheta_\mu(\xi,\epsilon)}Y_\mu^+,
\qquad
Y_\mu^{+\ddagger}\mapsto e^{-i\vartheta_\mu(\xi,\epsilon)}Y_\mu^{+\ddagger},
\]
one has
\[
Y_\mu^{+\ddagger}\partial_\xi Y_\mu^+
\mapsto
Y_\mu^{+\ddagger}\partial_\xi Y_\mu^+ + i\,\partial_\xi\vartheta_\mu.
\]
Hence choosing
\[
\partial_\xi\vartheta_\mu
=
-i\,Y_\mu^{+\ddagger}\partial_\xi Y_\mu^+
\]
enforces Eq.~\eqref{eq:app_adiabatic_parallel}.

For regular evanescent channels, there is no current-normalized phase condition of the same type, but one may still choose their amplitudes and phases continuously so that the corresponding wave-function vectors vary only at \(O(\delta\xi)\) across one parameter step.
Equivalently, after a smooth local scaling choice within each regular evanescent family, the full channel basis \(X(\xi,\epsilon)\) may be arranged so that
\begin{equation}
X(\xi+\delta\xi,\epsilon)-X(\xi,\epsilon)=O(\delta\xi)
\label{eq:app_regular_full_basis_variation}
\end{equation}
on every regular interval away from band edges.
This is the only property of the evanescent gauge needed below.

Thus, after these gauge and scaling choices, all regular channels vary smoothly with \(\xi\), and the connected propagating branches have no real \(O(\delta\xi)\) correction in their diagonal self-overlaps.

\subsection{Full-channel local matching away from band edges}
\label{appsubsec:adiabatic_away_edge}

We first consider a regular quasienergy at which no propagating branch approaches a band edge along the parameter step \(\xi_j\to\xi_{j+1}\).
In this case, all relevant propagating WFVs may be current normalized without singularity, and the full local basis varies as in Eq.~\eqref{eq:app_regular_full_basis_variation}.

Let
\[
X^{(j)}(\epsilon)\equiv X(\xi_j,\epsilon),
\qquad
X^{(j+1)}(\epsilon)\equiv X(\xi_{j+1},\epsilon)
\]
denote the full local channel bases on the two neighboring homogeneous segments.
The corresponding full local matching matrix is defined by
\begin{equation}
\mathcal M^{(j)}(\epsilon)
=
X^{(j)}(\epsilon)^\ddagger X^{(j+1)}(\epsilon).
\label{eq:app_full_matching_def}
\end{equation}
It maps channel amplitudes written in the basis of segment \(j+1\) to those written in the basis of segment \(j\).
Since the two local structures differ only by \(O(\delta\xi_j)\), and the basis has been chosen to vary smoothly with \(\xi\), one has
\begin{equation}
\mathcal M^{(j)}(\epsilon)=I+O(\delta\xi_j).
\label{eq:app_full_matching_regular}
\end{equation}

The connected propagating branch is obtained by identifying, inside this full-channel matching problem, the propagating mode that continues smoothly from segment \(j\) to segment \(j+1\).
For a regular propagating Bloch branch, the \(\xi\)-parallel transport gauge implies that the corresponding diagonal self-overlap has no real \(O(\delta\xi_j)\) correction.
Equivalently, the connected diagonal amplitude has the form
\begin{equation}
\mathcal M^{(j)}_{\conn,\conn}(\epsilon)
=
1+i\,O(\delta\xi_j)+O(\delta\xi_j^2),
\label{eq:app_connected_diag_amp}
\end{equation}
while the off-diagonal couplings to all other channels are \(O(\delta\xi_j)\).
Therefore the connected transmission \emph{probability} satisfies
\begin{equation}
\mathcal T_{\conn}^{(j)}(\epsilon)
=
1-O(\delta\xi_j^2)
\label{eq:app_local_regular_prob}
\end{equation}
for every regular propagating branch away from band edges.

This is the local statement used later in the long-boundary argument:
on a regular interval, the full-channel matching between neighboring local lattices preserves the connected propagating branch up to a probability loss of order \(O(\delta\xi_j^2)\).

\subsection{Band-edge partner block}
\label{appsubsec:adiabatic_partner}

We now consider the case where the parameter step approaches a band bottom or top.
Here the current of the relevant propagating branch tends to zero, so the current-normalized WFV becomes singular.
The local matching problem must therefore be reorganized in a partner basis containing both the propagating mode and its evanescent continuation.

Suppose two neighboring parameter values lie at distances \(O(\Delta_1)\) and \(O(\Delta_2)\) from a band-edge transition point, with
\[
\Delta_{1,2}<\delta\xi,
\qquad
\text{and at least one of }\Delta_1,\Delta_2\text{ is }O(\delta\xi)
\]
in the near-edge regime.
We assume the generic quadratic band-edge dispersion.
Then the current of a wavefunction-normalized partner state scales as
\begin{equation}
j \sim \sqrt{\Delta},
\label{eq:app_partner_current_scaling}
\end{equation}
so the corresponding current-normalization factor scales as
\begin{equation}
\mathcal N(\Delta)\sim \Delta^{-1/4}.
\label{eq:app_partner_norm_scaling}
\end{equation}

To describe the near-edge structure, choose a local two-dimensional partner frame \(\{Y_0,Y_{\mathrm{split}}\}\) which is smooth through the transition point.
Here \(Y_0\) is the neutral vector at the band edge, and \(Y_{\mathrm{split}}\) is the local splitting direction.
Then, to leading order in the distance from the edge, the two partner branches take the schematic form
\begin{equation}
Y_{\pm}^{\prop}(\Delta)
=
Y_0 \pm i\,c\,\Delta^{1/2}Y_{\mathrm{split}}+O(\Delta),
\qquad \Delta>0,
\label{eq:app_partner_prop_expansion}
\end{equation}
on the propagating side, and
\begin{equation}
Y_{\pm}^{\ev}(|\Delta|)
=
Y_0 \pm c\,|\Delta|^{1/2}Y_{\mathrm{split}}+O(|\Delta|),
\qquad \Delta<0,
\label{eq:app_partner_ev_expansion}
\end{equation}
on the evanescent side.
The same two-dimensional partner plane is therefore used on both sides of the transition; only the relative coefficient changes from real to purely imaginary.

We now write the full local matching matrix of Eq.~\eqref{eq:app_full_matching_def} in a basis where the distinguished row and column correspond to the propagating--evanescent partner channel crossing the band edge.
Factoring out the singular current normalizations, one may write
\begin{equation}
\mathcal M_{\edge}
=
D_1\,\widehat{\mathcal M}_{\edge}\,D_2,
\label{eq:app_edge_factorization}
\end{equation}
where
\begin{equation}
\begin{split}
D_1 &= \diag(1,\dots,1,O(\Delta_1^{-1/4}),1,\dots,1), \\
D_2 &= \diag(1,\dots,1,O(\Delta_2^{-1/4}),1,\dots,1).
\end{split}
\label{eq:app_edge_scalings}
\end{equation}

In this representation, the reduced matrix \(\widehat{\mathcal M}_{\edge}\) has the same structure as in the regular case away from the partner channel, while the central partner entry is controlled by the local partner expansions above.
More explicitly, its entries take the schematic form
\begin{equation}
\begin{split}
&\widehat{\mathcal M}_{\edge} = \\
&\left[
\begin{array}{@{}ccc|c|cc@{}}
\setlength{\arraycolsep}{2.5pt}
  1+\Im O(\delta \xi) & O(\delta\xi) & \cdots & O(\delta \xi) & \cdots & O(\delta\xi) \\
  O(\delta\xi) & 1+\Im O(\delta \xi) & \cdots & O(\delta \xi) & \cdots & O(\delta\xi) \\
  \vdots & \vdots & \ddots & \vdots & \ddots & \vdots \\
  \hline
  O(\delta \xi) & O(\delta \xi) & \cdots & \mathfrak m_{\edge} & \cdots & O(\delta \xi) \\
  \hline
  \vdots & \vdots & \ddots & \vdots & \ddots & \vdots \\
  O(\delta\xi) & O(\delta\xi) & \cdots & O(\delta \xi) & \cdots & 1+O(\delta \xi)
\end{array}
\right],
\end{split}
\label{eq:app_edge_matrix_raw}
\end{equation}
where the central partner scale satisfies
\begin{equation}
\mathfrak m_{\edge}
=
\begin{cases}
O(\Delta_1^{1/2})+O(\Delta_2^{1/2}), & \text{same-side step}, \\[4pt]
\sqrt{\,O(\Delta_1)+O(\Delta_2)\,}, & \text{crossing step}.
\end{cases}
\label{eq:app_edge_center_scale}
\end{equation}
In either case,
\begin{equation}
\mathfrak m_{\edge}=O(\delta\xi^{1/2})
\label{eq:app_edge_center_Osqrt}
\end{equation}
in the near-edge regime.

Introducing the auxiliary scaling
\begin{equation}
D_0=\diag(1,\dots,1,O(\delta\xi^{-1/4}),1,\dots,1),
\label{eq:app_edge_aux_scaling}
\end{equation}
one may equivalently rewrite the matching matrix as
\begin{equation}
\mathcal M_{\edge}
=
D_1D_0^{-1}
\Bigl[I+\mathcal E_{\edge}(\delta\xi)\Bigr]
D_0^{-1}D_2,
\label{eq:app_edge_regularized_block}
\end{equation}
where the rescaled correction \(\mathcal E_{\edge}(\delta\xi)\) has vanishing entries as \(\delta\xi\to0\).
Equivalently, after extracting the singular normalization factors, the partner block becomes regular and may be treated on the same footing as the regular full-channel matching problem.

The connected branch transmission is then obtained, exactly as in the regular case, only after solving this full-channel matching problem and identifying the branch that continues through the partner family.
Because the rescaled block is \(I+o(1)\), its inverse may be analyzed by the same Neumann-series logic as before.
For a propagating Bloch branch that continues through the band edge into its partner family, one again obtains
\begin{equation}
\mathcal T_{\conn}^{(j)}(\epsilon)
=
1-O(\delta\xi_j^2)
\label{eq:app_local_edge_prob}
\end{equation}
away from the measure-zero case of sitting exactly at the edge.

Thus the only effect of the band edge is that the local matching must be analyzed in a propagating--evanescent partner block, with the singular current normalization factored out explicitly.
After this is done, the connected branch still survives with asymptotically unit local transmission probability.

\subsection{Direct connected path along the long boundary}
\label{appsubsec:adiabatic_direct_path}

We now assemble the long adiabatic boundary from the local matching steps.
Fix one incoming open channel \(\nu\), and let the corresponding connected bulk branch family be followed through the discretized boundary.
Among all transmission paths across the \(N_s\) local interfaces, there is one distinguished \emph{direct connected path} \(s_*\) that follows this branch family throughout.

By Eqs.~\eqref{eq:app_local_regular_prob} and \eqref{eq:app_local_edge_prob}, every local step along the connected path has transmission probability
\[
1-O(\delta\xi_j^2).
\]
Therefore the total weight of the direct connected path is
\begin{equation}
\prod_{j=0}^{N_s-1}\bigl(1-O(\delta\xi_j^2)\bigr)
=
1-O\!\left(\sum_{j=0}^{N_s-1}\delta\xi_j^2\right).
\label{eq:app_direct_path_weight}
\end{equation}
For a partition with \(\max_j\delta\xi_j=O(N_s^{-1})\), this gives
\begin{equation}
1-O\!\left(\sum_{j=0}^{N_s-1}\delta\xi_j^2\right)
=
1-O(N_s^{-1}),
\label{eq:app_direct_path_weight_uniform}
\end{equation}
which tends to unity as \(N_s\to\infty\).

Thus the direct connected path already provides a lower bound approaching \(1\) for the averaged transmission into the connected bulk branch.
Combined with the shrinking-window dephasing argument used elsewhere in the paper, this yields the long-boundary adiabatic conclusion stated in the main text.

\section{Structure-preserving numerics, phase averaging, and Monte Carlo implementation}
\label{app:numerics}

This appendix collects the numerical ingredients used to evaluate the long-sample observables of the open Floquet problem.
The guiding principle is to mirror the analytic structure developed in the main text.
Accordingly, three features should be respected simultaneously:
frequency truncation should preserve Hermiticity of the frequency-space operator, spatial propagation should preserve the conjugate-symplectic structure of the transfer matrix, and long-sample observables should be evaluated through the same shrinking-window logic that underlies the analytic smoothing theory.

\subsection{Frequency truncation and Hermiticity preservation}
\label{appsubsec:numerics_truncation}

Let \(P_\Lambda\) be a finite-dimensional projector in frequency space, and define the truncated matrix
\begin{equation}
M_\Lambda(x,\epsilon)=P_\Lambda M(x,\epsilon)P_\Lambda.
\label{eq:app_num_MLambda}
\end{equation}
If the truncation is implemented as a genuine principal-subspace restriction of the Hermitian operator \(M\), then \(M_\Lambda\) remains Hermitian and the truncated generator
\begin{equation}
G_\Lambda(x,\epsilon)=
\begin{bmatrix}
0&I\\
M_\Lambda(x,\epsilon)&0
\end{bmatrix}
\label{eq:app_num_GLambda}
\end{equation}
still satisfies
\begin{equation}
G_\Lambda^\dagger J+JG_\Lambda=0.
\label{eq:app_num_inf_csp}
\end{equation}
Thus the infinitesimal conjugate-symplectic structure is preserved exactly at the truncated level.

The important point is not symmetry of the cutoff about \(n=0\) as such, but that the truncation be a genuine finite-dimensional restriction \(P_\Lambda M P_\Lambda\).
What matters is that Hermiticity be preserved, because this guarantees that the truncated transfer problem remains in the same conjugate-symplectic class as the full theory.

\subsection{Structure-preserving propagation}
\label{appsubsec:numerics_integrator}

After frequency truncation, one still needs a spatial propagation scheme that preserves the finite-step conjugate-symplectic identity.
Partition the sample into short intervals of length \(\Delta x\), centered at points \(x_j\), and approximate the generator on each interval by a local constant matrix \(G_\Lambda(x_j,\epsilon)\).
The exact local propagator is then
\begin{equation}
F_j(\epsilon)=\exp\!\bigl(G_\Lambda(x_j,\epsilon)\Delta x\bigr).
\label{eq:app_num_local_step}
\end{equation}
Because \(G_\Lambda^\dagger J+JG_\Lambda=0\), each local exponential satisfies
\begin{equation}
F_j(\epsilon)^\dagger JF_j(\epsilon)=J.
\label{eq:app_num_local_csp}
\end{equation}
Hence the product
\begin{equation}
F_\Lambda(x_M,x_0;\epsilon)=F_M(\epsilon)\cdots F_2(\epsilon)F_1(\epsilon)
\label{eq:app_num_product_step}
\end{equation}
also satisfies
\begin{equation}
F_\Lambda(x_M,x_0;\epsilon)^\dagger JF_\Lambda(x_M,x_0;\epsilon)=J,
\label{eq:app_num_global_csp}
\end{equation}
since the product of conjugate-symplectic matrices is again conjugate-symplectic.

This is the structure-preserving propagation scheme used here:
truncate in frequency so that \(M_\Lambda\) remains Hermitian, propagate by exact local exponentials, and multiply the resulting local propagators.
At the numerical level, this ensures that the algebraic backbone of the transfer-matrix formalism is respected up to floating-point roundoff rather than being broken already by the discretization itself.

\begin{remark}
More elaborate symplectic or Magnus-type schemes may certainly be used, but they are not conceptually required here.
The exact-exponential stepwise scheme already preserves the global conjugate-symplectic structure and is therefore well matched to the analytic framework of the paper.
\end{remark}

\subsection{From shrinking-window averaging to torus phase averages}
\label{appsubsec:phase_average}

The next task is to implement the long-sample observables.
In principle, one could evaluate the smoothed transmittance directly through the shrinking-window average
\[
\Delta\epsilon(L)\to0,
\qquad
L\Delta\epsilon(L)\to\infty,
\]
but for large \(L\) this quickly becomes inefficient because the transmission oscillates on energy scales of order \(L^{-1}\).
The same difficulty appears in branch-resolved quantities such as \(p_\mu\), since the underlying long-sample coefficients inherit the same rapidly varying propagation phases.

For fixed interfaces, the \(L\)-dependence of the finite-sample transmission matrix enters only through the propagating bulk phase factors
\[
D^\prop_\oplus(\epsilon)^L,
\qquad
D^\prop_\ominus(\epsilon)^{-L},
\]
or equivalently through the torus phase variables
\begin{equation}
\theta_j^+(\epsilon)=k_{\alpha_j}(\epsilon)L,
\qquad
\theta_j^-(\epsilon)=-k_{\beta_j}(\epsilon)L
\quad (\mathrm{mod}\ 2\pi),
\label{eq:app_num_phase_variables}
\end{equation}
assembled into
\[
\Theta(\epsilon;L)\in\mathbb T^{2N_B}.
\]
Thus every channel-resolved amplitude may be regarded, on a regular window, as a function
\begin{equation}
T_{\mu\nu}(\epsilon;L)
=
f_{\mu\nu}\!\left(
e^{i\theta_1^+},\dots,e^{i\theta_{N_B}^+};
e^{i\theta_1^-},\dots,e^{i\theta_{N_B}^-}
\right),
\label{eq:app_num_f_phase}
\end{equation}
with all rapid oscillations carried by the torus phase vector.

Assume that, on the shrinking window \(\mathcal E_L\) around a regular quasienergy \(\epsilon_0\), the phase velocities
\[
\partial_\epsilon \theta_j^\pm(\epsilon_0)
=
\pm L\,\partial_\epsilon k_j^\pm(\epsilon_0)
=
\pm \frac{L}{v_j^\pm(\epsilon_0)}
\]
are rationally independent after removing the common factor \(L\).
Equivalently, the inverse group velocities
\[
\left\{
\frac{1}{v_1^+(\epsilon_0)},\dots,\frac{1}{v_{N_B}^+(\epsilon_0)},
-\frac{1}{v_1^-(\epsilon_0)},\dots,-\frac{1}{v_{N_B}^-(\epsilon_0)}
\right\}
\]
satisfy no nontrivial integer relation.
This irrational-ratio condition is the generic case on a regular quasienergy window.
Under this condition, the map
\[
\epsilon\mapsto \Theta(\epsilon;L)
\]
becomes equidistributed on the full torus \(\mathbb T^{2N_B}\) as \(L\to\infty\).
Hence for any continuous torus function \(g(\Theta)\),
\begin{equation}
\lim_{L\to\infty}
\frac{1}{\Delta\epsilon(L)}
\int_{\mathcal E_L}
g\!\bigl(\Theta(\epsilon;L)\bigr)\,d\epsilon
=
\int_{\mathbb T^{2N_B}}
g(\Theta)\,\frac{d\Theta}{(2\pi)^{2N_B}}.
\label{eq:app_phase_average}
\end{equation}

Exactly the same logic applies to the branch-resolved open-sector quantities.
Once the deep-bulk coefficients are expressed as functions of the long propagation phases, the shrinking-window averages defining \(p_\mu\) and the channel-resolved population data \(p_{\mu\alpha}\) can be evaluated by the same phase-averaging strategy.

\subsection{Monte Carlo implementation}
\label{appsubsec:mc_estimator}

Once Eq.~\eqref{eq:app_phase_average} is available, the long-sample observables can be estimated by Monte Carlo sampling on the phase torus.
Let \(\Theta^{(s)}\), \(s=1,\dots,N_{\mathrm{MC}}\), be independent samples uniformly distributed on \(\mathbb T^{2N_B}\).
For the channel-resolved transmittance, the Monte Carlo estimator is
\begin{equation}
\widehat{\mathcal T}_{\mu\nu}^{(N_{\mathrm{MC}})}(\epsilon_0)
=
\frac{1}{N_{\mathrm{MC}}}
\sum_{s=1}^{N_{\mathrm{MC}}}
\mathcal T_{\mu\nu}\bigl(\epsilon_0;\Theta^{(s)}\bigr).
\label{eq:app_num_mc_estimator}
\end{equation}
Then:
\begin{enumerate}
\item the estimator is unbiased,
\[
\mathbb E\,
\widehat{\mathcal T}_{\mu\nu}^{(N_{\mathrm{MC}})}(\epsilon_0)
=
\bar{\mathcal T}_{\mu\nu}(\epsilon_0);
\]
\item if the variance is finite, then
\[
\mathrm{Var}\!\left[
\widehat{\mathcal T}_{\mu\nu}^{(N_{\mathrm{MC}})}(\epsilon_0)
\right]
=
\frac{\mathrm{Var}[\mathcal T_{\mu\nu}(\epsilon_0;\Theta)]}{N_{\mathrm{MC}}},
\]
so the root-mean-square statistical error scales as \(N_{\mathrm{MC}}^{-1/2}\);
\item the estimator converges almost surely to \(\bar{\mathcal T}_{\mu\nu}(\epsilon_0)\) by the strong law of large numbers.
\end{enumerate}

The same statements apply to other long-sample observables written as torus averages, including branch-resolved escape probabilities and channel-resolved branch-population data.

\subsection{Algorithmic summary}
\label{appsubsec:numerics_summary}

For clarity, we summarize the full numerical workflow used in the long-sample regime:
\begin{enumerate}
\item Choose a finite frequency projector \(P_\Lambda\) and form the truncated Hermitian matrix \(M_\Lambda=P_\Lambda M P_\Lambda\).
\item Build the truncated generator \(G_\Lambda\), which automatically satisfies the infinitesimal conjugate-symplectic identity.
\item Propagate in space using products of exact local exponentials \(\exp(G_\Lambda\Delta x)\), thereby preserving the finite-step conjugate-symplectic structure.
\item Extract the interface scattering matrices and construct the finite-sample transfer and transmission data.
\item In the long-sample regime, replace the shrinking-window energy smoothing by the phase average \eqref{eq:app_phase_average}.
\item Evaluate that phase average numerically by the Monte Carlo estimator \eqref{eq:app_num_mc_estimator}.
\end{enumerate}
This is the numerical counterpart of the analytic logic of the paper:
frequency truncation preserves the local generator structure, the propagation scheme preserves the global conjugate-symplectic structure, and phase averaging implements the same long-sample smoothing principle used in the theory.

\subsection{Why resonances slow convergence}
\label{appsubsec:mc_resonances}

In practice, the \(N_{\mathrm{MC}}^{-1/2}\) benchmark may be approached only very slowly when the transmission or branch-resolved observable develops ultra-narrow resonant structures on the phase torus.
Such structures do not invalidate the Monte Carlo formulation.
They instead increase the effective variance and thus delay the onset of the asymptotic statistical regime.
This is the natural numerical interpretation of the slow and irregular convergence seen in the deeper below-threshold cases.

\section*{Acknowledgments}
We are grateful to Fangcheng Wang for pointing out that the constrained form of the transfer matrix corresponds to a symplectic structure. We are also grateful to Carlo Beenakker for helpful discussions on MathOverflow regarding the structure of the symplectic group, and to Federico Poloni for pointing out relevant mathematical literature \cite{MOdiscussion,Mehrmann2016}. X.D. Dai acknowledges the support from the New Cornerstone Foundation. X. Dai is supported by the New Cornerstone Foundation and a fellowship and a CRF award from the Research Grants Council of the Hong Kong Special Administrative Region, China (Projects No. HKUST SRFS2324-6S01 and No. C7037-22GF).

\textit{Data Availability.}---  The source code and numerical scripts that support the findings of this study are available on GitHub \cite{Github_data}.

\bibliography{bib/others, bib/math, bib/wave_operator, bib/bound_state, bib/transport}

@article{Della2014,
  title = {Floquet-Hubbard bound states in the continuum},
  author = {Della Valle, Giuseppe and Longhi, Stefano},
  journal = {Phys. Rev. B},
  volume = {89},
  issue = {11},
  pages = {115118},
  numpages = {9},
  year = {2014},
  month = {Mar},
  publisher = {American Physical Society},
  doi = {10.1103/PhysRevB.89.115118},
  url = {https://link.aps.org/doi/10.1103/PhysRevB.89.115118}
}

@article{Yajima1983,
    author = {Yajima, Kenji and Kitada, Hitoshi},
    title = {Bound states and scattering states for time periodic hamiltonians},
    journal = {Annales de l'I.H.P. Physique th\'eorique},
    pages = {145--157},
    year = {1983},
    publisher = {Gauthier-Villars},
    volume = {39},
    number = {2},
    mrnumber = {722683},
    zbl = {0544.35073},
    url = {https://www.numdam.org/item/AIHPA_1983__39_2_145_0/}
}

@article{Kaneta1987, title={A class of time periodic Hamiltonians with no bound states}, volume={105}, DOI={10.1017/S0308210500021892}, number={1}, journal={Proceedings of the Royal Society of Edinburgh: Section A Mathematics}, author={Kaneta, H.}, year={1987}, pages={37–42}}

@article{Floquet1883,
  title = {Sur les équations différentielles linéaires à coefficients périodiques},
  volume = {12},
  ISSN = {1873-2151},
  url = {http://dx.doi.org/10.24033/asens.220},
  DOI = {10.24033/asens.220},
  journal = {Annales scientifiques de l’École normale supérieure},
  publisher = {Societe Mathematique de France},
  author = {Floquet,  G.},
  year = {1883},
  pages = {47–88}
}

@article{Shirley1965,
  title = {Solution of the Schr\"{o}dinger Equation with a Hamiltonian Periodic in Time},
  volume = {138},
  ISSN = {0031-899X},
  url = {http://dx.doi.org/10.1103/PhysRev.138.B979},
  DOI = {10.1103/physrev.138.b979},
  number = {4B},
  journal = {Physical Review},
  publisher = {American Physical Society (APS)},
  author = {Shirley,  Jon H.},
  year = {1965},
  month = may,
  pages = {B979–B987}
}

@article{Sambe1973,
  title = {Steady States and Quasienergies of a Quantum-Mechanical System in an Oscillating Field},
  volume = {7},
  ISSN = {0556-2791},
  url = {http://dx.doi.org/10.1103/PhysRevA.7.2203},
  DOI = {10.1103/physreva.7.2203},
  number = {6},
  journal = {Physical Review A},
  publisher = {American Physical Society (APS)},
  author = {Sambe,  Hideo},
  year = {1973},
  month = jun,
  pages = {2203–2213}
}

@book{Feng2010,
    author = {Kang ,Feng and Mengzhao, Qin},
    year = {2010},
    month = {10},
    title = {Symplectic Geometric Algorithms for Hamiltonian Systems},
    isbn = {978-7-5341-3595-8},
    doi = {10.1007/978-3-642-01777-3}
}

@book{Weyl1939,
  author = {Weyl, Hermann},
  title = {The Classical Groups: Their Invariants and Representations},
  publisher = {Princeton University Press},
  year = {1939},
  address = {Princeton, NJ},
}

@article{Pendry1994,
  title = {Symmetry and transport of waves in one-dimensional disordered systems},
  volume = {43},
  ISSN = {1460-6976},
  url = {http://dx.doi.org/10.1080/00018739400101515},
  DOI = {10.1080/00018739400101515},
  number = {4},
  journal = {Advances in Physics},
  publisher = {Informa UK Limited},
  author = {Pendry,  J.B.},
  year = {1994},
  month = aug,
  pages = {461–542}
}

@misc{Haber2017,
  title={What is the group of conjugate symplectic matrices},
  author={Haber, Howard E},
  year={2017},
  url={https://scipp-legacy.pbsci.ucsc.edu/~haber/webpage/conjuage_symplectic.pdf}
}

@article{Kitagawa2010,
  title = {Topological characterization of periodically driven quantum systems},
  author = {Kitagawa, Takuya and Berg, Erez and Rudner, Mark and Demler, Eugene},
  journal = {Phys. Rev. B},
  volume = {82},
  issue = {23},
  pages = {235114},
  numpages = {12},
  year = {2010},
  month = {Dec},
  publisher = {American Physical Society},
  doi = {10.1103/PhysRevB.82.235114},
  url = {https://link.aps.org/doi/10.1103/PhysRevB.82.235114}
}

@article{Rudner2013,
  title = {Anomalous Edge States and the Bulk-Edge Correspondence for Periodically Driven Two-Dimensional Systems},
  author = {Rudner, Mark S. and Lindner, Netanel H. and Berg, Erez and Levin, Michael},
  journal = {Phys. Rev. X},
  volume = {3},
  issue = {3},
  pages = {031005},
  numpages = {15},
  year = {2013},
  month = {Jul},
  publisher = {American Physical Society},
  doi = {10.1103/PhysRevX.3.031005},
  url = {https://link.aps.org/doi/10.1103/PhysRevX.3.031005}
}

@book{Born1999,
  title = {Principles of Optics: Electromagnetic Theory of Propagation,  Interference and Diffraction of Light},
  ISBN = {9781139644181},
  url = {http://dx.doi.org/10.1017/CBO9781139644181},
  DOI = {10.1017/cbo9781139644181},
  publisher = {Cambridge University Press},
  author = {Born,  Max and Wolf,  Emil and Bhatia,  A. B. and Clemmow,  P. C. and Gabor,  D. and Stokes,  A. R. and Taylor,  A. M. and Wayman,  P. A. and Wilcock,  W. L.},
  year = {1999},
  month = oct 
}

@article{Emmanouilidou2002,
  title = {Floquet scattering and classical-quantum correspondence in strong time-periodic fields},
  author = {Emmanouilidou, Agapi and Reichl, L. E.},
  journal = {Phys. Rev. A},
  volume = {65},
  issue = {3},
  pages = {033405},
  numpages = {15},
  year = {2002},
  month = {Feb},
  publisher = {American Physical Society},
  doi = {10.1103/PhysRevA.65.033405},
  url = {https://link.aps.org/doi/10.1103/PhysRevA.65.033405}
}

@article{Wenjun1999,
  title = {Floquet scattering through a time-periodic potential},
  author = {Li, Wenjun and Reichl, L. E.},
  journal = {Phys. Rev. B},
  volume = {60},
  issue = {23},
  pages = {15732--15741},
  numpages = {0},
  year = {1999},
  month = {Dec},
  publisher = {American Physical Society},
  doi = {10.1103/PhysRevB.60.15732},
  url = {https://link.aps.org/doi/10.1103/PhysRevB.60.15732}
}

@article{Martinez2001,
  title = {Transmission properties of the oscillating \ensuremath{\delta}-function potential},
  author = {Martinez, D. F. and Reichl, L. E.},
  journal = {Phys. Rev. B},
  volume = {64},
  issue = {24},
  pages = {245315},
  numpages = {9},
  year = {2001},
  month = {Dec},
  publisher = {American Physical Society},
  doi = {10.1103/PhysRevB.64.245315},
  url = {https://link.aps.org/doi/10.1103/PhysRevB.64.245315}
}

@article{Wigner1955,
  title = {Lower Limit for the Energy Derivative of the Scattering Phase Shift},
  author = {Wigner, Eugene P.},
  journal = {Phys. Rev.},
  volume = {98},
  issue = {1},
  pages = {145--147},
  numpages = {0},
  year = {1955},
  month = {Apr},
  publisher = {American Physical Society},
  doi = {10.1103/PhysRev.98.145},
  url = {https://link.aps.org/doi/10.1103/PhysRev.98.145}
}

@article{Smith1960,
  title = {Lifetime Matrix in Collision Theory},
  author = {Smith, Felix T.},
  journal = {Phys. Rev.},
  volume = {118},
  issue = {1},
  pages = {349--356},
  numpages = {0},
  year = {1960},
  month = {Apr},
  publisher = {American Physical Society},
  doi = {10.1103/PhysRev.118.349},
  url = {https://link.aps.org/doi/10.1103/PhysRev.118.349}
}

@article{Texier2016,
title = {Wigner time delay and related concepts: Application to transport in coherent conductors},
journal = {Physica E: Low-dimensional Systems and Nanostructures},
volume = {82},
pages = {16-33},
year = {2016},
note = {Frontiers in quantum electronic transport - In memory of Markus Büttiker},
issn = {1386-9477},
doi = {https://doi.org/10.1016/j.physe.2015.09.041},
url = {https://www.sciencedirect.com/science/article/pii/S1386947715302228},
author = {Christophe Texier},
}

@misc{Ljoint,
  title = {Boundary-Robust Transmission Asymmetry as a Topological Signature in Open Floquet Lattices (Joint Submit)},
  author = {Zhang, Ren and Ouyang, Xiao-yu and Dai, Xu-dong and Dai, Xi}
}

@misc{Github_data,
  author       = {Zhang, Ren and Ouyang, Xiao-Yu and Dai, Xu-Dong and Dai, Xi},
  title        = {Data and code for ``Scattering-state theory of open Floquet lattices: transfer matrices, branch openness, and robust asymmetry''},
  year         = {2026},
  howpublished = {\url{https://github.com/physiren/Data_Availability-Floquet_Scattering_States_Theory}},
  note         = {GitHub repository}
}

@misc{MOdiscussion,
    author = {Zhang, Ren},
    title = {Symplectic inner products of eigenvectors of complex symplectic matrix},
    howpublished = {MathOverflow},
    year = {2025},
    url = {https://mathoverflow.net/q/485629}
}

@article{Mehrmann2016,
    title = {On the sign characteristics of Hermitian matrix polynomials},
    journal = {Linear Algebra and its Applications},
    volume = {511},
    pages = {328-364},
    year = {2016},
    issn = {0024-3795},
    doi = {https://doi.org/10.1016/j.laa.2016.09.002},
    url = {https://www.sciencedirect.com/science/article/pii/S0024379516303913},
    author = {Volker Mehrmann and Vanni Noferini and Françoise Tisseur and Hongguo Xu},
}

@article{Thouless1983,
  title = {Quantization of particle transport},
  author = {Thouless, D. J.},
  journal = {Phys. Rev. B},
  volume = {27},
  issue = {10},
  pages = {6083--6087},
  numpages = {0},
  year = {1983},
  month = {May},
  publisher = {American Physical Society},
  doi = {10.1103/PhysRevB.27.6083},
  url = {https://link.aps.org/doi/10.1103/PhysRevB.27.6083}
}

@book{Newton2013,
  title = {Scattering Theory of Waves and Particles},
  author = {Newton, Roger G.},
  series = {Theoretical and Mathematical Physics},
  edition = {2},
  publisher = {Springer},
  address = {Berlin, Heidelberg},
  year = {2013},
  doi = {10.1007/978-3-642-88128-2},
  isbn = {978-3-642-88128-2},
  pages = {XX, 745},
  copyright = {Springer Science+Business Media New York 1982},
  note = {Originally published by McGraw Hill; Softcover ISBN: 978-3-642-88130-5 (Published: 18 April 2014); eBook ISBN: 978-3-642-88128-2 (Published: 27 November 2013)}
}

@article{Kato1966,
  title = {Wave operators and similarity for some non-selfadjoint operators},
  volume = {162},
  ISSN = {1432-1807},
  url = {http://dx.doi.org/10.1007/BF01360915},
  DOI = {10.1007/bf01360915},
  number = {2},
  journal = {Mathematische Annalen},
  publisher = {Springer Science and Business Media LLC},
  author = {Kato,  Tosio},
  year = {1966},
  month = jun,
  pages = {258–279}
}

@misc{Howland2012,
  doi          = {10.48550/ARXIV.1212.2931},
  url          = {https://arxiv.org/abs/1212.2931},
  author       = {Howland, J.},
  title        = {Scattering theory for Hamiltonians periodic in time},
  year         = {2012},
  howpublished = {arXiv:1212.2931},
}

@article{Howland1979,
    author = "James Howland",
     title = "Scattering Theory for Hamiltonians Periodic in Time",
   journal = "Indiana Univ. Math. J.",
  fjournal = "Indiana University Mathematics Journal",
    volume = 28,
      year = 1979,
     issue = 3,
     pages = "471--494",
      issn = "0022-2518",
     coden = "IUMJAB",
   mrclass = "",
}

\end{document}